\documentclass[aps, prd, floatfix, nofootinbib, superscriptaddress, twocolumn]{revtex4-2}

\usepackage{amsmath}
\usepackage{amssymb}
\usepackage{amsfonts}
\usepackage{color}
\usepackage{CJKutf8}
\usepackage{latexsym}
\usepackage{textcomp}

\usepackage[mathscr,scaled=1.15]{urwchancal}
\DeclareFontFamily{OT1}{pzc}{}
\DeclareFontShape{OT1}{pzc}{m}{it}%
{<-> s * [1.15] pzcmi7t}{}
\DeclareMathAlphabet{\mathpzc}{OT1}{pzc}{m}{it}

\usepackage{supertabular}
\usepackage{placeins}
\usepackage{epsfig}
\usepackage{graphicx}
\usepackage{multirow}

\definecolor{purple}{rgb}{0.5,0,0.5}
\definecolor{blue}{rgb}{0.0,0,0.9}
\definecolor{prdblue}{rgb}{0.133,0.118,0.498}
\usepackage[colorlinks=true, pdfstartview=FitV, linkcolor=prdblue, citecolor= prdblue, urlcolor=prdblue]{hyperref}

\newcounter{Afigure}
\newcounter{Atable}

\begin{document}

\begin{CJK}{UTF8}{gbsn}

\title{$\,$\\[-6.5ex]\hspace*{\fill}{\normalsize{\sf\emph{Preprint no}.\ USTC-ICTS/PCFT-26-86}}\\[1ex]
Quark + diquark description of nucleon gravitational form factors
}


\date{2026 November 12}  

\author{Peng Cheng%
    $\,^{\href{https://orcid.org/0000-0002-6410-9465}{\textcolor[rgb]{0.00,1.00,0.00}{\sf ID}}}$}
\affiliation{Department of Physics, \href{https://ror.org/05fsfvw79}{Anhui Normal University}, Wuhu, Anhui 241002, China}

\author{Zhen-Ni\ Xu%
$^{\href{https://orcid.org/0000-0002-9104-9680}{\textcolor[rgb]{0.00,1,0.00}{\sf ID}}}$}
\email[]{zhenni.xu@dci.uhu.es}
\affiliation{Dpto.\ Ciencias Integradas, Centro de Estudios Avanzados en Fis., Mat.\ y Comp.,
Fac.\ Ciencias Experimentales, \href{https://ror.org/03a1kt624}{Universidad de Huelva},
E-21071 Huelva, Spain}

\author{Dong-Yan Fu%
$^{\href{https://orcid.org/0000-0002-3205-9168}{\textcolor[rgb]{0.00,1,0.00}{\sf ID}}}$}
\affiliation{College of Physics, \href{https://ror.org/02wmsc916}{Guizhou University}, Guiyang 550025, China}

\author{Chen Chen%
$^{\href{https://orcid.org/0000-0003-3619-0670}{\textcolor[rgb]{0.00,1,0.00}{\sf ID}}}$}
\email[]{chenchen1031@ustc.edu.cn}
\affiliation{Interdisciplinary Center for Theoretical Study, \href{https://ror.org/04c4dkn09}{University of Science and Technology of China}, \\
Hefei, Anhui 230026, China}
\affiliation{\href{https://ror.org/01s9dmg43}{Peng Huanwu Center for Fundamental Theory}, Hefei, Anhui 230026, China}

\author{Zhao-Qian\ Yao%
$\,^{\href{https://orcid.org/0000-0002-9621-6994}{\textcolor[rgb]{0.00,1,0.00}{\sf ID}}}$}
\email{z.yao@hzdr.de}
\affiliation{\href{https://ror.org/01zy2cs03}{Helmholtz-Zentrum Dresden-Rossendorf},
Bautzner Landstra{\ss}e 400, D-01328 Dresden, Germany}

\begin{abstract}
We present parameter-free predictions for the nucleon gravitational form factors (GFFs) using a Poincar\'e-covariant quark+diquark Faddeev equation. We provide the full GFFs ($A$, $J$, $D$, and $\bar{c}$) along with their species decompositions into quark and gluon contributions. Based on the diquark breakdown, we find that while the scalar diquark component dominates the nucleon's energy, spin, and mechanical properties, the axial-vector component plays an indispensable role. The GFF $B(Q^2)$ is predicted to be vanishingly small but nonvanishing at finite $Q^2$; its physical origin is deeply rooted in the nucleon's internal structure through cancellations in both the diquark content and the $u/d$ flavor sectors. Furthermore, our results for $\bar{c}(Q^2)$ confirm the fundamental symmetry requirements of the approach. Compared with lQCD, our calculated $D$-term and various spatial radii are somewhat smaller in magnitude. This systematic underestimation is primarily driven by the gluon sector and can be attributed to the omission of meson cloud effects, suggesting that these effects impact the gluon distribution more significantly than the quark distribution. Finally, our analysis of the energy density based on two extrapolation schemes reveals that the presence of a negative central energy density is highly model-dependent.
\end{abstract}


\maketitle

\end{CJK}


%
\section{Introduction}

Understanding the nucleon as a bound system of three valence light quarks\,\cite{Hofstadter:1955ae} remains a central aim of ongoing experimental and theoretical efforts. The precise data on nucleon electromagnetic form factors, which map the spatial distributions of charge and magnetization, have been obtained from electron-nucleon scattering\,\cite{JeffersonLabHallA:1999epl, JeffersonLabHallA:2001qqe, Punjabi:2005wq, Cates:2011pz, Puckett:2017flj} and $e^+e^-$ annihilation\,\cite{BaBar:2013ves, BaBar:2013ukx, BESIII:2015axk, BESIII:2019hdp, BESIII:2021tbq}. The axial form factor, $G_A(Q^2)$, is determined from antineutrino-proton scattering, $\bar{\nu}_\mu + p \to \mu^+ + n$\,\cite{Ahrens:1988rr, Bodek:2007vi, Mosel:2016cwa, Meyer:2016oeg}, while the induced pseudoscalar form factor, $G_P(Q^2)$, is extracted from the longitudinal cross section in pion electroproduction\,\cite{Choi:1993vt, Bernard:1994pk, Fuchs:2003vw}. Although these observables are well-constrained experimentally and have been extensively studied theoretically\,\cite{Hohler:1976ax, Braun:2006hz, Cloet:2008re, Eichmann:2011vu, Bhattacharya:2013ehc, Chen:2023dxp, Lin:2021xrc, Yao:2024uej, Yan:2023nlb, Wang:2024abv, Cheng:2025yij, Alexandrou:2025vto, Jang:2019vkm, Chen:2021guo, Alexandrou:2021wzv, Jang:2023zts, Alexandrou:2023qbg, Chen:2024ksq, Salamu:2025cjv}, the nucleon gravitational form factors (GFFs), which encode the internal distributions of mass, spin, and mechanical properties, remain far less constrained.

Nucleon GFFs are defined from the matrix elements of the energy-momentum tensor (EMT). They encode fundamental properties such as the mass and spin of a state, the less well-known but equally fundamental $D$-term, and information interpretable as the distributions of energy, angular momentum, and mechanical stress\,\cite{Polyakov:2002yz, Polyakov:2018zvc, Lorce:2018egm, Burkert:2023wzr}. In particular, the mechanical stress distributions would be connected to the forces that confine quarks and gluons within a hadron, which is expected to be closely related to the QCD confinement mechanism\,\cite{Tanaka:2025pny}. Investigating these nonperturbative QCD properties connected to the internal stress distributions is therefore valuable for gaining new insights into the QCD confinement mechanism. Furthermore, because nucleon GFFs are weighted integrals over generalized parton distribution (GPD) functions, they are experimentally accessible via hard exclusive processes such as deeply virtual Compton scattering\,\cite{Ji:1996ek, Radyushkin:1997ki} and hard exclusive meson production\,\cite{Collins:1996fb}. 

On the theoretical side, nucleon GFFs have attracted increasing attention\,\cite{Chakrabarti:2020kdc, deTeramond:2021lxc, Amor-Quiroz:2023rke, Guo:2023qgu, Wang:2024abv, Cao:2024zlf, Broniowski:2025ctl} since the first experimental extraction in 2018\,\cite{Burkert:2018bqq}. A notable development is the recent lattice QCD (lQCD) calculation\,\cite{Hackett:2023rif} at a near-physical pion mass of  $m_\pi = 170\,\rm MeV$. However, the unphysical pion mass remains higher than the physical value, and the uncertainty of their results in $D$ form factor is relatively large near $Q^2 \approx 0$. Another important development is the model-independent determination of the pion and nucleon GFFs at the physical pion mass using a data-driven dispersive approach\,\cite{Cao:2024zlf}. In this approach, the $D$-term is determined to be $-3.38^{+0.34}_{-0.35}$, consistent with the lQCD result within uncertainties. Nonetheless, the resulting GFF uncertainties grow with $Q^2$, and no species decomposition is provided. Due to the systematic limitations of these methods, their predictions are currently restricted to $Q^2 \lesssim 2 \,\rm{GeV}^2$. Recently, a Poincar\'e-covariant three-body Faddeev equation approach was used to predict these form factors\,\cite{Yao:2024ixu}. However, owing to calculational challenges, it provides direct results only up to $Q^2 \lesssim 2.5\,\rm{GeV}^2$. Because calculating the spatial densities corresponding to GFFs via Fourier transformation requires complete knowledge of their $Q^2$-dependence, and extending to larger momentum transfers is crucial for connecting to perturbative QCD\,\cite{Tong:2021ctu, Tong:2022zax,Hatta:2025xuf}, an approach capable of providing predictions over a wider momentum range is highly desirable.

In this respect, the quark+diquark Faddeev equation approach\,\cite{Oettel:1998bk, Segovia:2015ufa} assumes that dynamical, nonpointlike diquark correlations play an important role in baryon structure. This assumption is rooted in a simplified treatment of the scattering problem within the two-body quark-quark subchannels of the full three-body problem (see, e.g., Ref.\,\cite{Hecht:2002ej}, Sect. II. A. 2). Such a simplification enables the approach to provide predictions for form factors at higher momentum transfers\,\cite{Cloet:2008re, Cheng:2025yij, Chen:2021guo, Segovia:2014aza, Segovia:2015hra, Lu:2019bjs}. Furthermore, this approach has been successfully used to describe and unify a wide array of properties of the nucleon, the $\Delta$-baryon, and their low-lying excitations, namely elastic and transition form factors\,\cite{Cloet:2008re, Cheng:2025yij, Chen:2021guo, Segovia:2014aza, Segovia:2015hra, Lu:2019bjs}. Ultimately, comparing the predictions of the three-body and quark+diquark approaches may reveal empirical signals for diquark correlations in nucleons.

In this work, we employ the quark+diquark Faddeev equation approach to calculate the proton GFFs, including their flavor decomposition into gluon, up-, down-, strange-, and charm-quark contributions. Based on these predicted GFFs, we extract the spatial distributions of energy, pressure, and shear force to explore the internal mechanical properties of the nucleon. Furthermore, we analyze the physical origin of the smallness of the GFF $B(Q^2)$, which has attracted recent attention\,\cite{Cao:2026jzm}. 

The remainder of this paper is organized as follows. Section\,\ref{sec:NGCurrents} is devoted to the description of the QCD EMT matrix elements and the definitions of the five nucleon GFFs. In Sect.\,\ref{sec:TheoFramework}, we briefly survey our theoretical framework, introducing the necessary elements for practical calculations of the GFFs. Section\,\ref{sec:resultsofGFFs} discusses our predictions for each nucleon GFF and compares them with available empirical inferences, existing lQCD results, and three-body approach predictions, with particular emphasis on the underlying mechanism for the smallness of $B(Q^2)$. The spatial density profiles, including energy, pressure, and shear force distributions, along with their corresponding radii, are described in Sect.\,\ref{sec:DensityandRadii}. Finally, Section\,\ref{sec:summary} presents a summary and perspective.

\section{Nucleon gravitational current}
\label{sec:NGCurrents}

The nucleon GFFs are defined via the matrix element of the EMT current\,\cite{Leader:2013jra,Burkert:2023wzr}
\begin{align}
&\hat{\Lambda}^{a}_{\mu \nu}(P_f,s_f;P_i,s_i) \nonumber \\
&= \left\langle P_f,s_f\left|T_{\mu \nu}^{a}(0)\right| P_i,s_i\right\rangle \nonumber \\
&= -\bar{u}(P_f,s_f) \tau^a \left[ P_{\mu} P_{\nu} A^a(Q^{2}) + i P_{\{\mu} \sigma_{\nu\} \rho} Q_{\rho} J^a(Q^2) \right. \nonumber \\
& \left. +\frac{1}{4} (Q_\mu Q_\nu - \delta_{\mu \nu} Q^2) D^a(Q^2) - i P_{[\mu} \sigma_{\nu] \rho} Q_{\rho} S^a(Q^2)\right. \nonumber \\
& \left. - \delta_{\mu \nu} m_N^2 \bar{c}^a(Q^2)
\right] u(P_i,s_i)/m_N\,,
\label{eq:EMTelements}
\end{align}
where $P_{i}$ ($P_{f}$) is the momentum of the incoming (outgoing) nucleon, with $P_{i,f}^2=-m_N^2$; $Q=P_{f} - P_{i}$, $P=(P_{i}+P_{f})/2$; all Dirac matrices are standard, with $\sigma_{\mu \nu}=(i/2)[\gamma_\mu,\gamma_\nu]$; $u(P,s)$ is the associated Euclidean spinor (see Appendix\,\ref{subapp:NAmplitude}); $a_{\{ \mu} b_{\nu \}}=(a_\mu b_\nu + a_\nu b_\mu)/2$ and $a_{[\mu} b_{\nu]}=(a_\mu b_\nu - a_\nu b_\mu)/2$. Throughout this paper, we work in the Euclidean space and use the $SU(2)_F$ isospin limit $m_u=m_d=:m_q$. The flavor structure is given by the Pauli matrices $\tau^a$: $\tau^0$ represents the total form factors and $\tau^{u/d}:=(\tau^0 \pm \tau^3)/2$ the valence u/d-quark contribution at hadron scale $\zeta_H$(see Sect.\,\ref{subsec:NGVertex} for details).

In Eq.\,\eqref{eq:EMTelements}\,\cite{Polyakov:2018zvc,Lorce:2017wkb}: $A^a$ is the nucleon momentum distribution form factor; $J^a$ relates to the nucleon spin distribution; $D^a$ provides information on the in-nucleon pressure and shear forces; and $S^a$ relates to the axial-vector form factor via $S^a(Q^2) =G_A^a(Q^2)/2$. The total form factors $\sum_{a} A^a$, $\sum_{a} J^a$, $\sum_{a} D^a$, $\sum_{a} \bar{c}^a$ are renormalization-scale invariant, whereas the individual quark and gluon form factors depend on the renormalization scale, which we omit for brevity. In the forward limit, $Q^2=0$, symmetries entail:
\begin{equation}
\label{eq:SymmetryConstrain}
\begin{aligned}
A(0)= \sum_{a} A^a(0) = 1,\ & J(0)= \sum_{a} J^a(0) = \frac{1}{2}\,, \\
S(0)= \sum_{a}\, S^a(0) = \frac{a_0}{2},\ & \bar{c}(0)= \sum_{a}\, \bar{c}^a(0) = 0\,,
\end{aligned}
\end{equation}
where $a_0$ is the singlet axial charge of the proton. The quantity $D(0)= \sum_{a} D^a(0) $ is also a conserved charge, but like the axial charge, $g_A$, its value is not constrained by any symmetry and must be determined dynamically. It has been described as the ``last unknown global property" of the nucleon\,\cite{Polyakov:2018zvc}. These form factors can be obtained directly from the current using an appropriate set of projection operators. For example, with 
\begin{equation}
\label{eq:GCurrent}
\begin{aligned}
&\Lambda^{a}_{\mu \nu}(P_f;P_i):=1/(4m_N^2) \\
&\ \times \sum_{s_f,s_i} u\left(P_f,s_f\right) \hat{\Lambda}^{a}_{\mu \nu}(P_f,s_f;P_i,s_i) \bar{u}\left(P_i,s_i\right),
\end{aligned}
\end{equation}
we define the following invariants: 
\begin{equation}
\label{eq:ProjectOperator}
\begin{array}{ll}
\mathpzc{s}_1^a = \frac{m_N}{P^{4}} \operatorname{Tr}[ P_{\mu} P_{\nu}\Lambda^{a}_{\mu \nu}]\,, &
\mathpzc{s}_2^a = \frac{i}{P^{2}} \operatorname{Tr}[ \gamma^{\perp}_{\mu} P_{\nu}\Lambda^{a}_{\mu \nu}]\,,
\\
\mathpzc{s}_3^a = \frac{m_N}{P^{2}} \operatorname{Tr}[ \delta_{\mu \nu}\Lambda^{a}_{\mu \nu}]\,, & \mathpzc{s}_4^a = \frac{m_N}{P^{2}} \operatorname{Tr}[ \gamma^{\perp}_{\mu} \gamma^{\perp}_{\nu}\Lambda^{a}_{\mu \nu}]\,, \\
\mathpzc{s}_5^a = \frac{i}{P^{2}} \operatorname{Tr}[ P_{\mu} \gamma^{\perp}_{\nu} \Lambda^{a}_{\mu \nu}]\,,
\end{array}
\end{equation}
where $Q\cdot \gamma^{\perp}=0$. The individual form factors are then extracted as
\begin{equation}
\label{eq:sistoGFFs}
\begin{array}{ll}
A^a = -\frac{5\mathpzc{s}_1^a + \mathpzc{s}_2^a - \mathpzc{s}_4^a + \mathpzc{s}_5^a}{4(1+\tau)}\,, &
J^a = -\frac{2\mathpzc{s}_1^a + \mathpzc{s}_2^a + \mathpzc{s}_5^a}{8 \tau}\,, \\
D^a = \frac{\mathpzc{s}_1^a + 2 \mathpzc{s}_3^a - 3 \mathpzc{s}_4^a}{4 \tau}\,, &
S^a =-\frac{\mathpzc{s}_2^a - \mathpzc{s}_5^a}{8 \tau}\,, \\
\bar{c}^a = \frac{\mathpzc{s}_4^a - \mathpzc{s}_3^a}{2}\,,
\end{array}
\end{equation}
with $\tau=Q^2/(4 m_N^2)$.

In addition to the GFFs discussed above, the GFF $B^a$ is also frequently discussed\,\cite{Polyakov:2018zvc, Cao:2026jzm, Lorce:2018egm}. Its relationship with the form factors $A^a$ and $J^a$ is as follows:
\begin{equation}
\label{eq:BGFF}
B^a(Q^2)=2 J^a(Q^2) -A^a(Q^2).
\end{equation}
Using the symmetry constraints from Eqs.\,\eqref{eq:SymmetryConstrain}, one obtains 
\begin{equation}
\label{eq:BGFF0}
B(0)=0,
\end{equation}
which is known as the vanishing of the ``anomalous gravitomagnetic moment" of a spin-$\frac{1}{2}$ fermion\,\cite{Kobzarev:1962wt}. 

\section{Theoretical framework}
\label{sec:TheoFramework}

\subsection{Quark+diquark Faddeev equation}
\label{subsec:FaddeevEquation}

\begin{figure}[t]
\centering
  \includegraphics[width=3.4in]{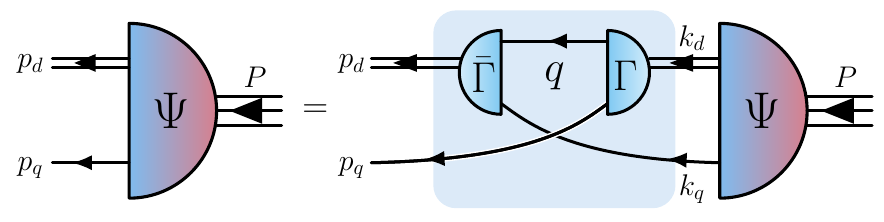}
\caption{\label{fig:FaddeevEquation}
The quark+diquark Faddeev equation, an integral equation for the Poincar\'e-covariant matrix-valued function $\Psi$, the Faddeev amplitude for a nucleon with total momentum $P=p_q+p_d=k_q+k_d$ constituted from three valence quarks, two of which are contained in a fully-interacting nonpointlike diquark correlation. $\Psi$ describes the relative momentum correlation between the dressed quarks and diquarks.
Legend. \emph{Shaded rectangle} -- Faddeev kernel;
\emph{single line} -- dressed-quark propagator [App.\,\ref{subapp:qprop}];
$\Gamma$ -- diquark correlation amplitude [App.\,\ref{subapp:diqAmp}] and \emph{double line} -- diquark propagator[App.\,\ref{subapp:diqprop}]. Only isoscalar-scalar, $[q_1 q_2]$, and isovector-axialvector, $\{q_1 q_2\}$, diquarks play a role in nucleons.}
\end{figure}

The diquark correlations simplify analyses of the three-valence-quark scattering problem, and hence, baryon bound states, because they reduce this task to solving a Poincar\'e-covariant Faddeev equation\,\cite{Cahill:1988dx,Efimov:1990uz,Oettel:1999gc}, as depicted in Fig.\,\ref{fig:FaddeevEquation}.

In the quark+diquark picture, the binding of a nucleon (and kindred baryons) arises from two contributions: one from the formation of tight (but not pointlike) diquark correlations, and the other from the attraction between the quark and diquark generated through quark exchange (shaded area in Fig.\,\ref{fig:FaddeevEquation}). This exchange ensures that diquark correlations within the baryon are fully dynamical: no quark holds a special place because each one participates in all diquarks to the fullest extent allowed by its quantum numbers. The continual rearrangement of the quarks guarantees, among other
things, that the nucleon’s dressed-quark wave function satisfies Pauli statistics. 

We now specify the elements of the Faddeev equation kernel: the dressed-quark propagator $S(p)$, the diquark propagators $\Delta^{J^P}(P)$, and the Bethe-Salpeter (BS) amplitudes $\Gamma^{J^P}(p;P)$. The quark+diquark framework here uses an efficient algebraic parametrization for the dressed light-quark propagator, encodes confinement and dynamical chiral symmetry breaking, retains the leading diquark amplitudes, and describes diquark propagation consistently with color confinement and asymptotic freedom (see Appendix\,\ref{app:qdiqframework}). It has recently provided a wide range of predictions for various baryons, including their spectra\,\cite{Chen:2017pse, Chen:2019fzn, Liu:2022ndb, Liu:2022nku} and form factors\,\cite{Cloet:2008re, Segovia:2014aza, Segovia:2015hra, Segovia:2016zyc, Chen:2018nsg, Lu:2019bjs, Cheng:2025yij}. 

It is worth stressing that the Faddeev equation describes the nucleon's dressed-quark core; all statements here therefore refer to this part of the nucleon. Solving the nucleon's Faddeev equation yields a core mass of $m_N=1.18\,\rm GeV$. The complete (physical) nucleon is obtained by including resonant (meson-cloud) contributions to the Faddeev kernel, which are known to 
produce a reduction of about $0.2\,\rm GeV$ in the physical nucleon mass\,\cite{Sanchis-Alepuz:2014wea}. While the impact of such effects on nucleon structure can be estimated using dynamical coupled-channels models\,\cite{Aznauryan:2012ba}, their explicit inclusion lies beyond the scope of contemporary Faddeev equation analyses.

\subsection{Baryon-Graviton Vertex}
\label{subsec:NGVertex}

To compute the GFFs, one must specify how the graviton couples to the constituents of the composite hadrons. In the present context, this amounts to specifying the couplings of the graviton to the dressed quarks and the diquark correlations, given that the incoming and outgoing nucleons are described by the quark-diquark Faddeev amplitudes. 

\begin{figure}[t]
  \centering
  \includegraphics[width=3.4in]{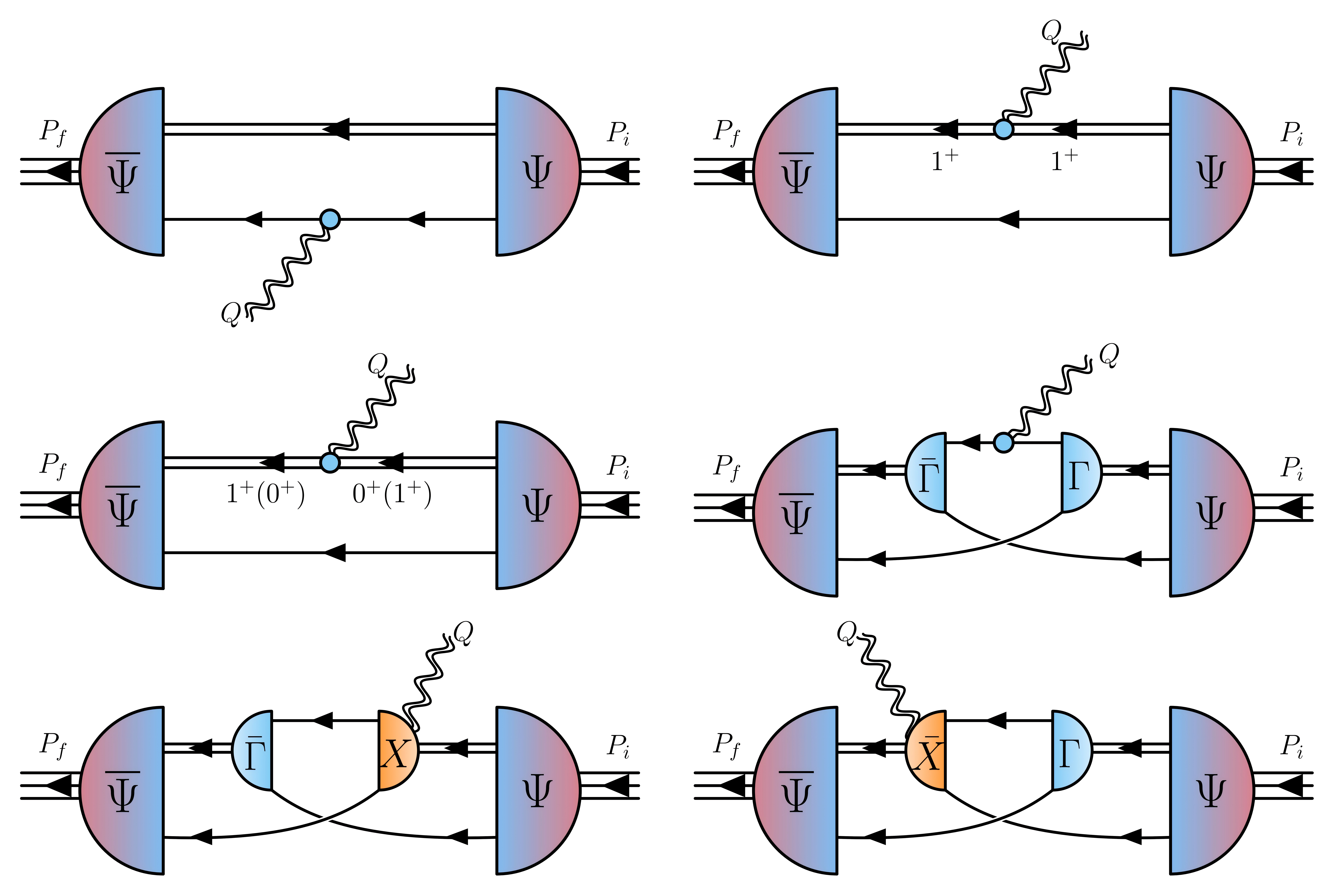}
  \caption{\label{fig:NGVertex} The collection of gravitational interaction diagrams that ensures a conserved current for on-shell nucleons described by the Faddeev amplitudes, $\Psi$, produced by the equation depicted in Fig.\,\ref{fig:FaddeevEquation}. Legend: \emph{single line}, dressed quark propagator; \emph{double undulating line}, gravitational probe; $\Gamma$,  diquark correlation amplitude; \emph{double line}, diquark propagator; and $X$, seagull terms. The top-left image is Diagram 1; the top-right, Diagram 2; and so on, with the bottom-right image, Diagram 6.}
\end{figure}

Following the systematic procedure of Ref.\,\cite{Oettel:1999gc}, we construct the graviton-nucleon vertex $\hat{\Lambda}^{a}_{\mu \nu}(P_f;P_i)$ in Eq.\,\eqref{eq:EMTelements}. This scheme automatically yields a conserved current for on-shell nucleons described by the Faddeev amplitudes calculated above. We stress that the current in Fig.\,\ref{fig:NGVertex} is the minimal form required for current conservation: each contribution is necessary, and their sum guarantees the outcome without overcounting or undercounting. This construction follows from the requirement that the graviton couples to every quark in the nucleon described by the Faddeev equation in Fig.\,\ref{fig:FaddeevEquation}. By starting from the Faddeev equation and applying the minimal substitution rule, we identify the individual contributions: Fig.\,\ref{fig:NGVertex}(1) is the leading contribution; Fig.\,\ref{fig:NGVertex}(2, 3) express the graviton penetrating to connect with the quarks in the diquark; Fig.\,\ref{fig:NGVertex}(4) represents the graviton striking the exchanged quark; and Fig.\,\ref{fig:NGVertex}(5, 6) are the “seagull” terms arising from the momentum dependence of the diquark BS amplitude. While the derivation in Ref.\,\cite{Oettel:1999gc} is general, it has been extended to the axialvector current in Ref.\,\cite{Chen:2021guo} and refined in the electromagnetic current\,\cite{Cheng:2025yij}; here, we extend it to the gravitational current. Canonical normalization of the nucleon Faddeev amplitude is implemented by fixing the proton Dirac form factor $F_1(0) = 1$ \,\cite{Cheng:2025yij}.

The interaction current has six terms, which are depicted in Fig.\,\ref{fig:NGVertex} and represent the following expression: 
\begin{equation}
\label{eq:NGravitonVertex}
\begin{aligned}
& \hat{\Lambda}^{a}_{\mu \nu}(P_f,s_f;P_i,s_i) = \int \frac{d^4 p}{(2 \pi)^4} \frac{d^4 k}{(2 \pi)^4} \bar{u}(P_f,s_f) \\
&\sum_{\hat{i}=1}^6 \bar{\Psi}\left(p, -P_f\right) J_{\mu \nu}^{\hat{i},a}\left(p, P_f ; k, P_i\right) \Psi(k, P_i) u(P_i,s_i),
\end{aligned}
\end{equation}
where $\bar{\Psi}\left(p, -P_f\right)=C^\dagger \Psi\left(-p, -P_f\right)^T C$. In practical calculations, we work in the Breit frame: $P_{f/i}=P^{BF} \pm Q/2$, $P^{BF}=(0,0,0,i E)$, $E^2=m_N^2+Q^2/4$.  

It is worth noting that our approach is formulated at the hadron scale, $\zeta_H$, where all properties of a hadron are carried by its valence degrees of freedom. Consequently, partonic glue and sea-quark contributions to observable form factors are absorbed into the dressed valence quarks at $\zeta_H$. This is merely a definition of dressing, as reflected in any complete QCD Schwinger function (see, e.g., the diagrammatic expansion of the quark gap equation).
 
As noted above, the hadron scale $\zeta_H$ plays a central role in our approach. Its existence and character are discussed in Ref.\,\cite{Yin:2023dbw}, and its utility has been proven in developing predictions for various parton distribution and fragmentation functions\,\cite{Ding:2022ows,Raya:2021zrz,dePaula:2022pcb,Xu:2024nzp,Xu:2025cyj,Xing:2023pms,Xing:2025eip}. Because the total nucleon GFFs are scale-invariant, formulating the calculations at $\zeta_H$ has no effect on them, but it does affect their decomposition into contributions from different parton species. An example is provided in Ref.\,\cite{Yao:2024ixu}. Using the charge discussed in Refs.\,\cite{Cui:2019dwv,Deur:2023dzc}, the hadron scale is predicted to be: $\zeta_H=0.331(2)\,\rm{GeV}$\,\cite{Cui:2021mom}. With the framework thus specified, we now turn to the individual contributions in Eq.\,\eqref{eq:NGravitonVertex}.

\subsubsection{Diagram~1}
\label{subsubsec:Diag1}

This diagram expresses the graviton coupling directly to the bystander quark: 
\begin{align}
\label{eq:diagram1}
J_{\mu \nu}^{\hat{1},a}\left(p, P_f;k, P_i\right) & = i \chi_{\mu \nu}^{q,a}(p_q, k_q) \left(\Delta^{0^+}(k_d) + \Delta^{1^+}(k_d) \right) \nonumber \\
& \times (2\pi)^4 \delta^4(p-k-\hat{\eta}Q)\,,
\end{align}
where, for notational simplicity, we have suppressed the subscript on $\Delta^{1^+}$, Eq.\,\eqref{eq:diqprop}, which connect to partners in the Faddeev amplitude; $\chi^{q,a}_{\mu \nu}(p_q,k_q):=\tau^a \chi^{q}_{\mu \nu}(p_q,k_q)$
is the unamputated dressed-graviton-quark vertex; and the momenta are as follows($\eta + \hat{\eta}=1$):
\begin{eqnarray}
\label{etavalue}
\begin{array}{lc@{\qquad}l}
k_q=\eta P_i+k\,, &  p_q=\eta P_f+p\,, \\
k_d=\hat{\eta}P_i-k\,, &  p_d=\hat{\eta}P_f-p\,.
\end{array}
\end{eqnarray}

We use $\eta=1/3$ throughout, so that the incoming graviton momentum is shared equally among the dressed valence quarks in the nucleon. This simplifies numerical analyses and accelerates the convergence of calculations. Naturally, because our approach is Poincar\'e covariant, the results are independent of $\eta$ so long as the numerical methods preserve this covariance.

Current conservation requires that the unamputated graviton-quark vertex satisfies a Ward–Green–Takahashi identity (WGTI):
\begin{equation}
\label{eq:GWGTI}
i (\ell_1 - \ell_2)_\mu \, \chi^{q}_{\mu \nu}(\ell_1,\ell_2) = \ell_{2 \nu} S(\ell_2) -S(\ell_1) \ell_{1 \nu}  \,.
\end{equation}
 Since the dressed quark propagator $S(p)$ has been parameterized, we draw on Ref.\,\cite{Cheng:2025yij} and use the following Ansatz for the vertex, expressed entirely in terms of quantities already specified in Eq.\,\eqref{eq:qpropagator}: 
\begin{align}
\label{eq:qgvertex}
i \chi^{q}_{\mu \nu}\left(\ell_1, \ell_2\right) & = \ell_\nu \Big[i\gamma_\mu \Sigma_{\sigma_V} + 2\ell_\mu \left[i \gamma \cdot \ell  \Delta_{\sigma_V} - \Delta_{\sigma_S}\right] \nonumber \\
 & + \frac{i}{2}[s_1-\bar{s}_1]\left[\gamma \cdot Q \gamma_\mu \gamma \cdot \ell- \gamma \cdot \ell \gamma_\mu \gamma \cdot Q\right] \Delta_{\sigma_V}  \nonumber \\
& - i [s_2-\bar{s}_2] \sigma_{\mu \rho} Q_\rho  \Delta_{\sigma_S} \nonumber \\
& -\frac{1}{2} [1+s_3] \gamma \cdot Q  \sigma_{\mu \rho} Q_\rho \Delta_{\sigma_V}\Big] \nonumber \\
& - \frac{\delta_{\mu \nu}}{2} \left[S(\ell_1) + S(\ell_2)\right] + i \chi^{q T}_{\mu \nu}(\ell_1, \ell_2), 
\end{align}
with 
\begin{equation}
\begin{aligned}
s_i & =a_i+b_i \exp [-\frac{1}{4}\mathcal{E}(Q) / M_q^E], \\
\bar{s}_{1,2}&=1-s_{1,2}, \\
\mathcal{E}(Q) / M_q^E & =\left(1 + Q^2 / [2 M^E_q]^2\right)^{1 / 2}-1,
\end{aligned}
\label{eq:Momentumfactor2}
\end{equation}
where $Q=\ell_1-\ell_2$, $\ell=[\ell_1+\ell_2]/2$, $M^E_q=0.33\,\rm GeV$ (shown in Appendix\,\ref{subapp:qprop}), and 
\begin{equation}
\begin{aligned}
\Sigma_F= &
\frac{1}{2} [F(\ell_1^2)+F(\ell_2^2)], \\
\Delta_F= &
[F(\ell_1^2)-F(\ell_2^2)]/[\ell_1^2-\ell_2^2].    
\end{aligned}
\label{eq:DefofFun}
\end{equation}
Here, the parameters $a_{1, 2, 3}$ and $b_{1, 2, 3}$ characterize the transverse part of the vertex Ansatz. They have been determined in our previous study of the electromagnetic form factor\,\cite{Cheng:2025yij} and are listed in Table\,\ref{Tab:Parameters}.

\begin{table}[t]
\caption{\label{Tab:Parameters} Parameters determined in the study of the electromagnetic form factors\,\cite{Cheng:2025yij}.
(The radii are in units of fm.)}
\begin{center}
\begin{tabular*}
{\hsize}
{
c@{\extracolsep{0ptplus1fil}}
c@{\extracolsep{0ptplus1fil}}
c@{\extracolsep{0ptplus1fil}}
c@{\extracolsep{0ptplus1fil}}
c@{\extracolsep{0ptplus1fil}}
c@{\extracolsep{0ptplus1fil}}
c@{\extracolsep{0ptplus1fil}}
c@{\extracolsep{0ptplus1fil}}
c@{\extracolsep{0ptplus1fil}}
c@{\extracolsep{0ptplus1fil}}}\hline
 $r^{em}_{\rm sc}$
  & $r^{em}_{\rm av}$ &   $a_1$ & $a_2$ & $a_3$ & $b_1$ & $b_2$ & $b_3$\\\hline
 $0.31$ & $1.05$ & $0.17$ & $1.47$ & $-0.93$ & $1.03$ & $-0.17$ & $1.61$ \\\hline
\end{tabular*}
\end{center}
\end{table}

The remaining piece in Eq.\,\eqref{eq:qgvertex}, $\chi^{q T}_{\mu \nu}(\ell_1, \ell_2)$, satisfies $Q_\mu \chi^{q T}_{\mu \nu}(\ell_1, \ell_2) = 0$. It represents all remaining transverse structures not already included in the previous terms and does not contribute to resolving Eq.\,\eqref{eq:GWGTI}. As stressed in Ref.\,\cite{Xu:2023izo}, isoscalar scalar mesons emerge in the graviton-quark vertex, leading to the following term for the lightest scalar meson:
\begin{equation}
\chi^{q T,sc}_{\mu \nu}(\ell_1, \ell_2)=i S(\ell_1)T_{\mu \nu}(Q) S(\ell_2) P_q^{\mathbb{S}}\left(Q^2\right),
\label{eq:Scalarterm}
\end{equation}
where 
\begin{equation}
P_q^{\mathbb{S}}(t=Q^2)=\frac{-t\,r_{\mathbb{S}_q}}{t+m_{\mathbb{S} q}^2},
\label{eq:Scalarterm1}
\end{equation}
with the predicted $m_{\mathbb{S}_q}=0.55\,\rm{GeV}$ and $r_{\mathbb{S}_q}=0.193\,\rm{GeV}$(which translates into $f_{\mathbb{S}_q}=0.026\,\rm{GeV}$)\,\cite{Xu:2023izo}. Since the term in Eq.\,\eqref{eq:Scalarterm} plays a dominant role among all other transverse structures, we use: 
\begin{equation}
\chi^{q T}_{\mu \nu}(\ell_1, \ell_2)= i S(\ell_1)T_{\mu \nu}(Q) S(\ell_2) P_q^{\mathbb{S}}\left(Q^2\right).
\label{eq:gtrans}
\end{equation}
We have checked other terms, such as tensor meson-pole terms, and found that they have almost no influence on the results.

\subsubsection{Diagram~2}
\label{subsubsec:Diag2}

In this case, the graviton probes the scalar or axialvector diquark:
\begin{equation}
\label{eq:diagram2}
\begin{aligned}
J_{\mu \nu}^{\hat{2},a}\left(p, P_f ; k, P_i\right) & = \left[\chi_{\mu \nu}^{dq,a}(p_{d}, k_{d}) \right] S(k_q) \\
& \times (2\pi)^4 \delta^4(p-k+\eta Q)\,,    
\end{aligned}
\end{equation}
where
\begin{equation}
 [\chi_{\mu \nu}^{dq,a}(p_{d};k_{d})]={\rm diag}[
F^{0^+, a} \chi_{\mu \nu}^{0^+}, F^{1^+, a}\chi_{\mu \nu}^{1^+}], 
\label{eq:Gdiquarkvertex}
\end{equation}
with $\chi_{\mu \nu}^{0^+}$ given in Eq.\,\eqref{eq:chi0plus}, $\chi_{\mu \nu}^{1^+}$ given in Eq.\,\eqref{eq:chi1plus}, and the color-flavor coefficients $F^{0^+(1^+), a}$ obtained from Eq.\,\eqref{eq:cfsofDia2&3}.

If one is dealing with an elementary scalar correlation, then the WGTI of the unamputated graviton-scalar-diquark vertex reads:
\begin{equation}
\label{eq:dqsWTI}
Q_\mu \,\chi_{\mu \nu}^{0^+}(\ell_1,\ell_2) = \ell_{2\nu} \Delta^{0^+}(\ell_2) -
 \Delta^{0^+}(\ell_1) \ell_{1\nu}.
\end{equation}
However, for a composite system of the type we consider, this identity is modified, similar to the case in the electromagnetic current \cite{Segovia:2014aza}:
\begin{equation}
\label{eq:dqsWTIComp}
Q_\mu \,\chi_{\mu \nu}^{0^+}(\ell_1,\ell_2) = \left[\ell_{2\nu} \Delta^{0^+}(\ell_2) -
 \Delta^{0^+}(\ell_1) \ell_{1\nu}\right] F_{\rm J=sc}(Q^2)\,,
\end{equation}
where
\begin{equation}
\label{eq:Fdqsform}
F_{\rm J}(Q^2) = \frac{1}{1 + \frac{1}{6} r_{\rm J}^2 Q^2}
\end{equation}
is a form factor describing the distribution of mass within the correlation.

We employ the following Ansatz for the unamputated graviton-scalar-diquark vertex:
\begin{equation}
\label{eq:chi0plus}
\begin{aligned}
\chi_{\mu \nu}^{0^+}(\ell_1,\ell_2) & = -\big[ 2 \ell_\mu \ell_\nu\,
\Delta_{\Pi^{0^+}}(\ell_1^2,\ell_2^2)\, \\
& + \delta_{\mu \nu}\Sigma_{\Pi^{0^+}}(\ell_1^2,\ell_2^2) \big]\,F_{\rm sc}(Q^2)\,,    
\end{aligned}
\end{equation}
where $\Delta_{F}$ and  $\Sigma_{F}$ are defined in Eq.\,\eqref{eq:DefofFun}, $\Pi^{J^P}$ is defined in Eq.\,\eqref{eq:defofPiJP},  
$\ell=[\ell_1+\ell_2]/2$ and the value of $r_{sc}$ is determined in Eqs.\,\eqref{eq:radius}. This Ansatz satisfies Eq.\,\eqref{eq:dqsWTIComp} and is completely determined by quantities already introduced; moreover, it is free of kinematic singularities in the relevant domain.

For the axial-vector diquark, the WGTI of the unamputated graviton–axialvector-diquark vertex reads:
\begin{equation}
Q_\mu \chi_{\mu\nu,\alpha\beta}^{1^{+}}\left(\ell_1, \ell_2\right)=\left[\ell_{2\nu} \Delta_{\alpha\beta}^{1^{+}}\left(\ell_2\right) -\Delta_{\alpha\beta}^{1^{+}}\left(\ell_1\right) \ell_{1\nu}\right]F_{\rm av}(Q^2).
\label{eq:dqaxWTIComp}
\end{equation}
We use the following Ansatze for this unamputated vertex, expressed completely in terms of the functions that characterize the diquark propagators in Eqs.\,\eqref{eq:diqprop}: 
\begin{align}
\chi_{\mu\nu,\alpha\beta}^{1^{+}}\left(\ell_1, \ell_2\right)& = \Big[-\ell_{\nu}\big[2 \delta_{\alpha\beta}\ell_\mu \Delta_{\Pi^{1^+}}(\ell_1^2, \ell_2^2) \nonumber \\
& +\frac{1} {m_{1^+}^2} \left(\delta_{\mu\alpha}\ell_{1\beta}\Pi^{1^+}\left(\ell_1^2\right)+\delta_{\mu\beta}\ell_{2\alpha}\Pi^{1^+}\left(\ell_2^2 \right)\right) \nonumber \\
& +\frac{2}{m_{1^+}^2}\ell_{2\alpha}\ell_{1\beta}\ell_\mu \Delta_{\Pi^{1^+}}(\ell_1^2, \ell_2^2)\big] \nonumber \\
&  - \delta_{\mu \nu}\frac{\Delta_{\alpha\beta}^{1^{+}}\left(\ell_1\right)+\Delta_{\alpha\beta}^{1^{+}}\left(\ell_2\right)}{2}\Big]F_{av}(Q^2),  
\label{eq:chi1plus}
\end{align}
with the value of $r_{\rm av}$ determined in Eqs.\,\eqref{eq:radius}. 

To fix the remaining parameters in Eqs.\,\eqref{eq:chi0plus} and \eqref{eq:chi1plus}, we follow Ref.\,\cite{Xu:2023izo}, which predicts the ratio of the pion mass radius to its charge radius to be $r_{\pi}^{\theta_2}/r_{\pi}^F = 0.74$, consistent with the value $0.79(3)$ from recent data analyses\,\cite{Xu:2023bwv}. Given the connection between mesons and diquarks\,\cite{Barabanov:2020jvn,Segovia:2014aza}, we use 
\begin{equation}
\begin{aligned}
r_{\rm{sc}}=0.74\,r_{\rm{sc}}^{em} = 0.23\,\rm{fm}, \\
r_{\rm{av}}=0.74\,r_{\rm{av}}^{em}=0.78\,\rm{fm},   
\end{aligned}
\label{eq:radius}
\end{equation}
where $r_{\rm sc}^{em}$ and $r_{\rm av}^{em}$ are given in Table\,\ref{Tab:Parameters}. From Eq.\,\eqref{eq:Fdqsform}, the result is only weakly sensitive to the values of these two parameters.

\subsubsection{Diagram~3}
\label{subsubsec:Diag3}
This image depicts the form factor contribution from gravitationally induced $0^+ \leftrightarrow 1^+$ diquark transitions: 
\begin{align}
\label{eq:diagram3}
J_{\mu \nu}^{\hat{3},a}\left(p, P_f ; k, P_i\right) &= 
\Delta^i(p_{d})\left[ \Gamma_{\mu \nu}^{dq,a}(p_{d},k_{d}) \right]^{\{i \neq j\}}\Delta^j(k_{d}) \nonumber \\
& \times S(k_q) (2\pi)^4 \delta^4(p-k+\eta Q)\,,
\end{align}
where $[\Gamma_{\mu \nu}^{dq,a}(p_{d},k_{d})]^{\{01\}}=F^{01,a}\,\Gamma^{01}_{\mu \nu \alpha}(p_{d},k_{d})$ and
$[\Gamma_{\mu \nu}^{dq,a}(p_{d};k_{d})]^{\{10\}}=F^{10,a}\,\Gamma^{10}_{\mu \nu \alpha}(p_{d},k_{d})$, with the color-flavor coefficients $F^{01(10), a}$ obtained from Eq.\,\eqref{eq:cfsofDia2&3}. This transition vertex is a rank-3 pseudotensor and can therefore be expressed as
\begin{align}
\label{SAGVertex}
\Gamma^{01}_{\mu \nu \alpha}(\ell_1,\ell_2) & =
-\Gamma^{10}_{\mu \nu \alpha}(\ell_1,\ell_2) \nonumber \\
&= \frac{i}{m_N} \, {\cal T}_{sa}(\ell_1,\ell_2) \,
\varepsilon_{\mu\alpha\rho\lambda}\ell_{1\rho} \ell_{2 \lambda} \ell_{\nu}\,,
\end{align}
where $\mu$ and $\nu$ are the tensor indices of the graviton, and $\alpha$ is the vector index of the axial-vector diquark. As in the electromagnetic current, we assume 
\begin{equation}
\label{calTvalue}
{\cal T}_{sa}(\ell_1,\ell_2) = \kappa_{sa} F_{sa}(Q^2) \,.
\end{equation}
We take $\kappa_{sa} = 2.64$ and $r_{sa} = r_{av}$, the same as in the electromagnetic current\,\cite{Cheng:2025yij}. Note that since $F^{0 1}=F^{1 0}=0$, this diagram does not contribute to the total form factors, but it does affect the flavor decomposition.

\subsubsection{Diagram~4}
\label{subsubsec:Diag4}

This two-loop diagram represents the form factor contribution produced when the graviton couples to the exchanged quark as one diquark breaks up and another is formed. It is expressed as follows:
\begin{align}
\label{eq:diagram4}
J_{\mu\nu}^{\hat{4},a}\left(p, P_f ; k, P_i\right) & = -\frac{1}{2} S(p_{q}) \Delta^{i}(p_{d}) \Gamma^{j}(p_1;k_{d}) i \chi_{\mu \nu}^{q,a \rm T}(q',q) \nonumber \\
& \times  \bar{\Gamma}^{i \rm T}(p'_2;-p_{d})
\Delta^j(k_{d}) S(k_{q})\,,    
\end{align}
where the graviton-quark vertex, $\chi_{\mu \nu}^{q,a}$, appears in Eq.\,\eqref{eq:diagram1} and the charge-conjugated correlation amplitude is defined after Eqs.\,\eqref{eq:Amplitudenorm}. While this is the first two-loop diagram we have described, no new elements appear in its specification: the dressed graviton-quark vertex was discussed in Sect.\,\ref{subsubsec:Diag1}. In Eq.\,\eqref{eq:diagram4}, the momenta are
\begin{eqnarray}
\begin{array}{lc@{\qquad}l}
q = \hat{\eta}P_i-\eta P_f-p-k\,, & & q' = \hat{\eta}P_f-\eta P_i-p-k \,,\\
p_1 = (p_q-q)/2\,,& & p'_2 = (-k_q+q')/2 \,,\\
p'_1 = (p_q-q')/2\,, & & p_2 = (-k_q+q)/2 \,.
\end{array}
\end{eqnarray}
The full contribution (four terms) is obtained by summing over the superscripts $\{i= 0^+$, $1^+; j=0^+$, $1^+\}$. When required, the axialvector diquark subscripts can readily be made explicit.

Notably, quark exchange provides the attraction necessary
to bind the nucleon in the Faddeev equation. It also ensures that the Faddeev amplitude has the correct antisymmetry under the exchange of any two dressed quarks. This key feature is absent in models with elementary (non-composite) diquarks.  

\subsubsection{Diagrams~5 and 6}
\label{subsubsec:Diag5&6}
These two-loop diagrams are typically called ``seagull'' terms, which appear as partners of Diagram~4 and arise because binding in the nucleon Faddeev equation is generated by the exchange of a dressed-quark between \textit{nonpointlike} diquark correlations\,\cite{Oettel:1999gc}. The explicit expressions for their contribution to the nucleon form factors are
\begin{align}
\label{eq:diagram5}
J_{\mu \nu}^{\hat{5},a}\left(p, P_f ; k, P_i\right) &= \frac{1}{2} S(p_{q}) \Delta^{i}(p_{d}) X_{\mu \nu}^{j,a}(p_1';k_d) S^{\rm T}(q') \nonumber \\
& \times \bar{\Gamma}^{i \rm T}(p'_2;-p_{d}) \Delta^{j}(k_{d}) S(k_{q})\,,    
\end{align}
for Diagram 5, and 
\begin{align}
\label{eq:diagram6}
J_{\mu \nu}^{\hat{6},a}\left(p, P_f ; k, P_i\right) &=- \frac{1}{2} S(p_{q}) \Delta^{i}(p_{d}) \Gamma^{j}(p_1;k_{d}) S^{\rm T}(q) \nonumber \\
& \times \bar{X}_{\mu \nu}^{i,a}(p_2;-p_d) \Delta^{j}(k_{d}) S(k_{q})\,,
\end{align}
for Diagram 6; here, again, the superscripts are summed, and, when necessary, the axialvector diquark subscripts can be made explicit.

The new elements here are the couplings of a graviton to two dressed quarks as they either separate from (Diagram~5) or combine to form (Diagram~6) a diquark correlation. They are therefore components of the five-point Schwinger function which describes the coupling of a graviton to the quark-quark scattering kernel. This Schwinger function can be calculated, as demonstrated by the computation of analogous Schwinger functions relevant to meson observables\,\cite{Cotanch:2002vj,Ding:2019lwe}. 

However, such a calculation would provide relevant input only when a
uniform truncation of the Dyson-Schwinger equations is employed to calculate each of the elements described hitherto. Absent that, we instead employ an algebraic parametrization, inspired by Refs.\,\cite{Oettel:1999gc,Chen:2021guo}, which for Diagram~5, Eq.\,\eqref{eq:diagram5}, reads
\begin{align}
& X^{J^P,a}_{\mu \nu}(\ell;K) \nonumber \\
& = \tau^a p_{q\nu} \frac{4 \ell_\mu- Q_\mu}{4 \ell\cdot Q -
Q^2}\,\left[\Gamma^{J^P}\!(\ell-Q/2)-\Gamma^{J^P}\!(\ell)\right] \nonumber \\
& + \frac{4 \ell_\mu+ Q_\mu}{4 \ell\cdot Q +
Q^2}\,\left[\Gamma^{J^P}\!(\ell+Q/2)-\Gamma^{J^P}\!(\ell)\right] q_\nu \tau^{a \rm T} \nonumber \\
& + \frac{\delta_{\mu \nu}}{2} (\tau^a \Gamma^{J^P}(\ell) + \Gamma^{J^P}(\ell) \tau^{a \rm T}), 
\label{eq:seagull}    
\end{align}
with $\ell$ the relative momentum between the quarks in the initial diquark, $K$ the total momentum of the initial diquark, $p_q$ the momentum of the quark which becomes the bystander, and $q$ the momentum of the quark reabsorbed into the final diquark. Diagram~6, Eq.\,\eqref{eq:diagram6}, has
\begin{align}
& \bar X^{J^P,a}_{\mu \nu}(\ell; -K) \nonumber \\
& = \frac{4 \ell_\mu+ Q_\mu}{4 \ell\cdot Q +
Q^2}\,\left[\bar\Gamma^{J^P}\!(\ell+Q/2)-\bar\Gamma^{J^P}\!(\ell)\right] p_{q \nu} \tau^a\nonumber \\
& + \tau^{a \rm T} q_\nu \frac{4 \ell_\mu-Q_\mu}{4 \ell\cdot Q -
Q^2}\,\left[\bar\Gamma^{J^P}\!(\ell-Q/2)-\bar\Gamma^{J^P}\!(\ell)\right] \nonumber \\
& - \frac{\delta_{\mu \nu}}{2} (\bar\Gamma^{J^P}\!(\ell) \tau^a +\tau^{a \rm T} \bar\Gamma^{J^P}\!(\ell)), 
\label{eq:barseagull}
\end{align}
where $\bar\Gamma^{J^P}\!(\ell)$ is the charge-conjugated amplitude. 

It is worth stressing that, like Eq.\,\eqref{eq:chi0plus}, Eqs.\,\eqref{eq:seagull} and \eqref{eq:barseagull} are simple forms, free of kinematic singularities and sufficient to ensure the nucleon-graviton interaction vertex (Fig.\,\ref{fig:NGVertex}) satisfies the relevant WGTI when the composite nucleon is obtained from the Faddeev equation sketched in Fig.\,\ref{fig:FaddeevEquation}. Thus, they are sufficient for our purposes, as in Refs.\,\cite{Chen:2021guo,Cheng:2025yij}.

\section{Gravitational form factors: results}
\label{sec:resultsofGFFs}

\subsection{Numerical results at hadron scale $\zeta_H$}
\label{subsec:PredictionatHS}

To compute all form factors, we use the algorithm
described in Appendix D of Ref.\,\cite{Segovia:2014aza}. For the Faddeev equation, ten Chebyshev polynomials are employed to represent the $\ell \cdot P$ dependence of the Faddeev amplitudes $\Psi^{0^+/1^+}(\ell; P)$ in Eqs.\,\eqref{eq:AplitudeBasis}. For the integrals depicted in Fig.\,\ref{fig:NGVertex}, we use $150$ quadrature points for the momentum magnitude and $50$ for each angular variable for Diagrams $1–3$, and $1.5 \times 10^8$ Monte Carlo points for Diagrams $4–6$. These choices ensure stable results for $Q^2 \lesssim 10\, \rm{GeV}^2$.

Having established the numerical framework, we will subsequently discuss the $Q^2$-dependence of the nucleon GFFs and also present contribution separations according to the diquark breakdown:

\hspace*{1em}\parbox[b]{0.9\linewidth}{(\emph{\rm total}) -- sum of all contributions;\\
(\emph{s}) -- sum of all contributions involving a scalar diquark in both the incoming and outgoing nucleon $\Psi_{i,f}$;\\
(\emph{a}) -- sum of all contributions involving an axialvector diquark in both the incoming and outgoing nucleon $\Psi_{i,f}$;\\
(\emph{m}) -- sum of all contributions in which the diquark component of $\Psi_i$ differs from that in $\Psi_f$.}
\smallskip 

\begin{figure}[t]
\leftline{\hspace*{0.5em}{\large{\textsf{A}}}}
\vspace*{-5ex}
\includegraphics[clip, width=0.42\textwidth]{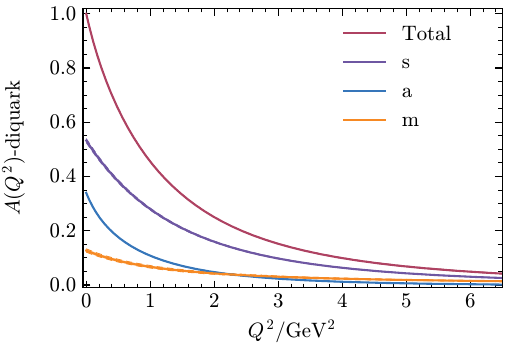}
\vspace*{4ex}

\leftline{\hspace*{0.5em}{\large{\textsf{B}}}}
\vspace*{-5ex}
\includegraphics[clip, width=0.42\textwidth]{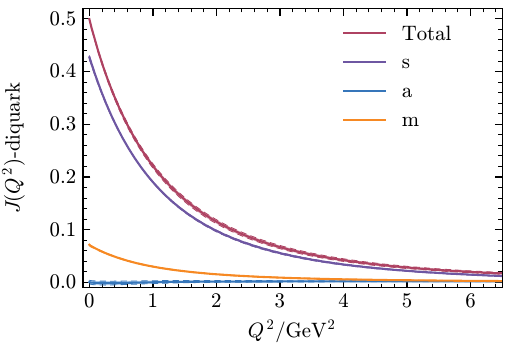}
\vspace*{4ex}

\leftline{\hspace*{0.5em}{\large{\textsf{C}}}}
\vspace*{-5ex}
\includegraphics[clip, width=0.42\textwidth]{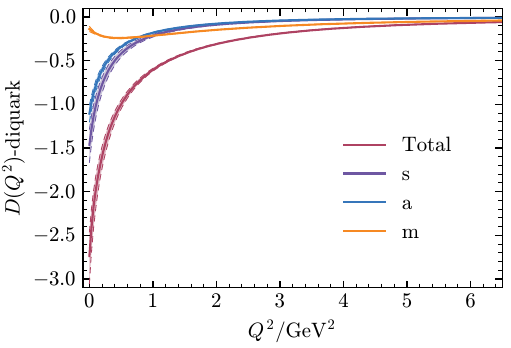}
\caption{\label{fig:protonGFFs} Proton gravitational form factors $A(Q^2)$, $J(Q^2)$ and $D(Q^2)$ (the uncertainty bands arise from variations of the diquark mass): the full result and its decomposition according to the diquark content of the diagrams.}
\end{figure}

\begin{figure}[t]
\leftline{\hspace*{0.5em}{\large{\textsf{A}}}}
\vspace*{-5ex}
\includegraphics[clip, width=0.42\textwidth]{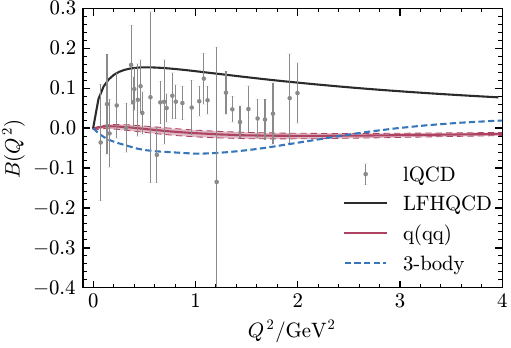}
\vspace*{4ex}

\leftline{\hspace*{0.5em}{\large{\textsf{B}}}}
\vspace*{-5ex}
\includegraphics[clip, width=0.42\textwidth]{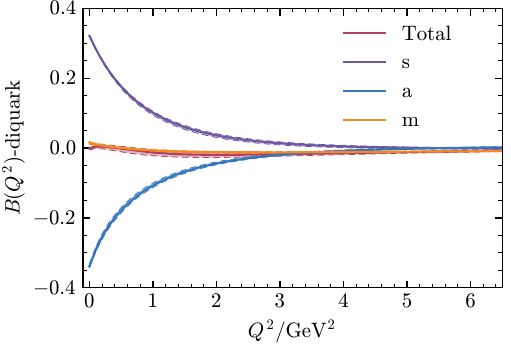}
\vspace*{4ex}

\leftline{\hspace*{0.5em}{\large{\textsf{C}}}}
\vspace*{-5ex}
\includegraphics[clip, width=0.42\textwidth]{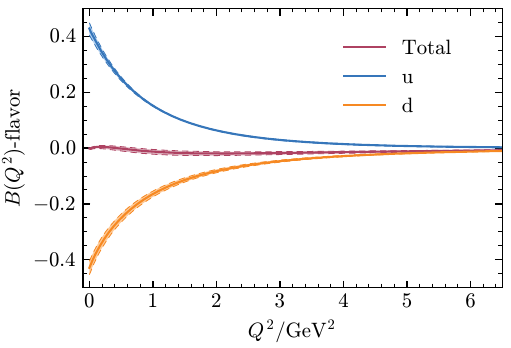}
\caption{\label{fig:BGFFs} Proton gravitational form factors $B(Q^2)$ (uncertainty bands as in Fig.\,\ref{fig:protonGFFs}). A: the full result compared with results obtained from lQCD (dots\,\cite{Hackett:2023rif}), the three-body approach (dashed line\,\cite{Yao:2024ixu}), and LFHQCD (solid line\,\cite{Cao:2026jzm}). B: the full result and its decomposition according to the diquark content of the diagrams; C: the full result and its flavor decomposition.}
\end{figure}

Our results for the proton GFFs $A(Q^2)$, $J(Q^2)$ and $D(Q^2)$ are shown in Fig.\,\ref{fig:protonGFFs}. The light-colored bands represent the impact of $\pm 5\%$ variations in the diquark masses, with the parameters in Table\,\ref{Tab:Parameters} varied by $\mp 3\%$ so as to preserve the description of the electromagnetic form factors obtained in our previous work\,\cite{Cheng:2025yij}. Focusing on $Q^2=0$ and working with Fig.\,\ref{fig:protonGFFs}A, one can define a set of momentum contributions. The scalar diquark component contributes $P_A^s=A^s(Q^2=0)=53.3^{+0.7}_{-0.7}\%$; the axialvector diquark component contributes $P_A^a=A^a(Q^2=0)=34.0^{-0.1}_{+0.0} \%$; and their interference contributes $P_A^m=A^m(Q^2=0)=12.7^{-0.6}_{+0.7}\%$. \textit{N.B.} The upper and lower uncertainties correspond to $+5\%$ and $-5\%$ variations of the diquark mass, respectively; the signs of the shifts follow from the actual numerical results. These observations are close to the separation obtained from the canonical normalization\,\cite{Liu:2022ndb}. Similarly, Figure\,\ref{fig:protonGFFs}B reveals the corresponding spin contributions: the scalar diquark component contributes $85.5^{+0.9}_{-0.6}\%$ to the proton spin, the interference between the scalar and axial-vector diquark components contributes $14.3^{-0.2}_{+0.0}\%$, and the axialvector diquark component contributes only $0.2^{-0.7}_{+0.6}\%$, indicating a negligible contribution to the proton spin. Clearly, for the GFF $J(0)$, the scalar diquark component plays a dominant role. Our results for the proton GFF $D(Q^2)$ are shown in Fig.\,\ref{fig:protonGFFs}C. Its value at $Q^2=0$, known as the $D$-term, has been described as the “last unknown global property” of the nucleon. Figure\,\ref{fig:protonGFFs}C reveals the following contributions to the $D$-term: the scalar diquark component contributes $54.4^{+1.2}_{-1.2} \%$, the axialvector diquark component accounts for $40.4^{+0.3}_{-0.5}\%$, and the interference between the scalar and axialvector diquark components provides $5.2^{-1.5}_{+1.7}\%$. These results indicate that both the scalar and axialvector diquark components make significant contributions to the $D$-term.

Figure\,\ref{fig:BGFFs}A shows our results of the GFF $B(Q^2)$. Our results predict that the magnitude of $B(Q^2)$ is vanishingly small but nonvanishing at finite $Q^2$ and that it strictly obeys the constraint $B(0)=0$, in agreement with the three-body approach\,\cite{Yao:2024ixu} and the light-front holographic QCD (LFHQCD)\,\cite{Cao:2026jzm}. Figure\,\ref{fig:BGFFs}B reveals that the underlying mechanism for the smallness of $B(Q^2)$ is the mutual cancellation between the scalar and axialvector diquark components. Figure\,\ref{fig:BGFFs}C indicates that the physical origin of this smallness can also be understood as a flavor cancellation between the $u$ and $d$ quarks. Throughout this analysis, we have shown that the smallness of $B(Q^2)$ is not accidental but is deeply rooted in the nucleon's internal structure, as also supported by Ref.\,\cite{Cao:2026jzm}. 

\begin{figure}[t]
\centering
  \includegraphics[clip, width=0.46\textwidth]{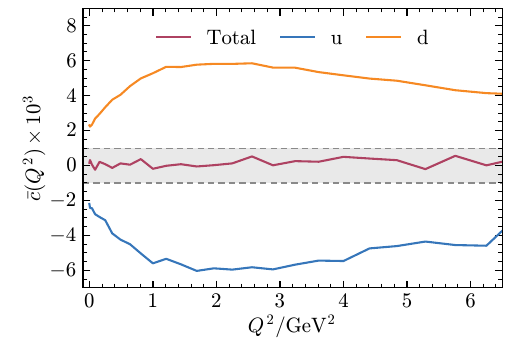}
\caption{\label{fig:cGFFs} Full result for the proton gravitational form factor $\bar{c}(Q^2)$ and its flavor decomposition. (Fluctuations reflect the numerical precision of our calculation.)}
\end{figure}

Our numerical result for $\bar{c}(Q^2)$, shown in Fig.\,\ref{fig:cGFFs}, indicates that the computed $\bar{c}(Q^2)$ is essentially indistinguishable from zero over the entire range of momentum transfer shown. This confirms that our approach satisfies the fundamental symmetry requirements without any fine tuning; rather, it follows automatically from the careful construction of the currents discussed above. Figure\,\ref{fig:cGFFs} further shows that $\bar{c}^{u/d}(Q^2)$ is nonvanishing but extremely small at the hadron scale $\zeta_H$. Consequently, these contributions can be neglected in the analysis of the flavor-decomposed energy and pressure densities inside the proton\,\cite{Lorce:2018egm}. In addition, our result for the GFF $S$ at $Q^2=0$ is $S(0)=0.401^{+0.014}_{-0.018}$, which is somewhat larger than the value of $0.325(10)$ derived from the singlet axial charge within the same framework\,\cite{Chen:2022odn}.

\subsection{Predictions at $\zeta=2\,\rm GeV$}
\label{subsec:Predat2GeV}

\begin{table}[t]
\caption{\label{TableZero}
$Q^2=0$ values of proton gravitational form factors and species decompositions at $\zeta = \zeta_2 = 2$\,GeV, including a charm ($c$) quark sea contribution. \textit{N.B.} The upper and lower uncertainties correspond to a $+5\%$ and $-5\%$ variation of the diquark mass, respectively; for entries where the uncertainty is smaller than the displayed precision, it is omitted.
The lQCD and three-body results are reproduced from Refs.\,\cite{Hackett:2023rif} and \cite{Yao:2024ixu}, respectively, where the former reports no signal for a $c$ contribution. 
Throughout the table, $q$ denotes the sum over all quark contributions(valence and sea).}
\begin{tabular*}
{\hsize}
{
l@{\extracolsep{0ptplus1fil}}
|l@{\extracolsep{0ptplus1fil}}
l@{\extracolsep{0ptplus1fil}}
l@{\extracolsep{0ptplus1fil}}}\hline\hline
herein$\ $ & $A(0)\ $ & $J(0)\ $ & $-D(0)\ $ \\\hline
Total$\ $ & $1.00\ $ & $0.500^{+0.001}_{-0.001}\ $ & $2.736^{+0.319}_{-0.270}\ $ \\
$u\ $ & $0.360\ $ & $0.282^{+0.006}_{-0.006}\ $ & $0.909^{+0.131}_{-0.106}\ $ \\
$d\ $ & $0.191\ $ & $-0.0069^{-0.005}_{+0.005}\ $ & $0.598^{+0.045}_{-0.043}\ $ \\
$s\ $ & $0.025\ $ & $0.012\ $ & $0.067^{+0.008}_{-0.007}\ $ \\
$c\ $ & $0.0085\ $ & $0.0042\ $ & $0.023^{+0.003}_{-0.002}\ $ \\
$q\ $ & $0.584\ $ & $0.292^{+0.001}_{-0.001}\ $ & $1.598^{+0.186}_{-0.158}\ $ \\
$g\ $ & $0.416\ $ & $0.208$ & $1.138^{+0.133}_{-0.112}\ $ \\
\hline\hline
three-body$\ $ & $A(0)\ $ & $J(0)\ $ & $-D(0)\ $ \\\hline
Total$\ $ & $1.00\ $ & $0.50\ $ & $3.114(10)_\pm\ $ \\
$u\ $ & $0.361(15)_\pm\ $ & $0.181(08)_\pm\ $ & $1.120(50)_\pm\ $ \\
$d\ $ & $0.191(08)_\pm\ $ & $0.095(04)_\pm\ $ & $0.594(25)_\pm\ $ \\
$s\ $ & $0.025(02)_\mp\ $ & $0.012(02)_\mp\ $ & $0.077(06)_\mp\ $ \\
$c\ $ & $0.0086(05)_\mp\ $ & $0.0043(03)_\mp\ $ & $0.027(02)_\mp\ $ \\
$q\ $ & $0.584(13)_\pm\ $ & $0.292(06)_\pm\ $ & $1.820(43)_\pm\ $ \\
$g\ $ & $0.416(13)_\mp\ $ & $0.208(06)_\mp\ $ & $1.294(33)_\mp\ $ \\
\hline\hline
lQCD$\ $ & $A(0)\ $ & $J(0)\ $ & $-D(0)\ $ \\\hline
Total$\ $ & $1.011(37)\ $ & $0.506(25)\ $ & $3.87(97)\ $ \\
$u\ $ & $0.3255(92)\ $ & $0.2213(85)\ $ & $0.56(17)\ $ \\
$d\ $ & $0.1590(92)\ $ & $0.0197(85)\ $ & $0.57(17)\ $ \\
$s\ $ & $0.0257(95)\ $ & $0.0097(82)\ $ & $0.18(17)\ $ \\
$q\ $ & $0.510(25)\ $ & $0.251(21)\ $ & $1.30(49)\ $ \\
$g\ $ & $0.501(27)\ $ & $0.255(13)\ $ & $2.57(84)\ $ \\\hline\hline
\end{tabular*}
\end{table}

Although the total form factors are scale-invariant, the individual quark and gluon form factors are scale-dependent. To compare with the lQCD results, we evolve the above results from the hadron scale $\zeta_H$ to $\zeta= 2\,\rm GeV$. This is enabled by the all-orders (AO) evolution scheme detailed in Ref.\,\cite{Yin:2023dbw}, whose key principles we summarize here. (a) There is an effective charge, $\alpha_{1\ell}(k^2)$, in the sense of Refs.\,\cite{Grunberg:1980ja,Grunberg:1982fw} and as reviewed in Ref.\,\cite{Deur:2023dzc}. When used to integrate the leading-order perturbative DGLAP equations, it defines an evolution scheme for every parton distribution function (DF) that is all-orders exact. The process-independent strong running coupling deﬁned and computed in Ref.\,\cite{Cui:2019dwv} possesses all requisite properties. (b) There is a scale, $\zeta_H$, at which all properties of a given hadron are carried by its valence degrees of freedom. 

\begin{figure}[t]
\leftline{\hspace*{0.5em}{\large{\textsf{A}}}}
\vspace*{-5ex}
\includegraphics[clip, width=0.42\textwidth]{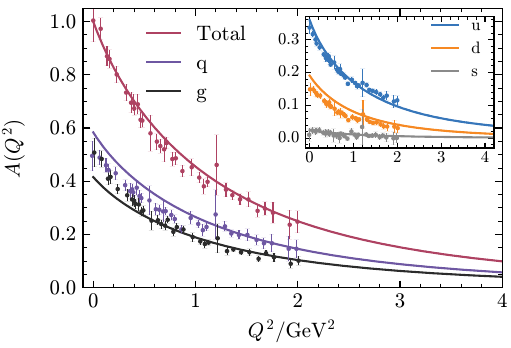}
\vspace*{4ex}

\leftline{\hspace*{0.5em}{\large{\textsf{B}}}}
\vspace*{-5ex}
\includegraphics[clip, width=0.42\textwidth]{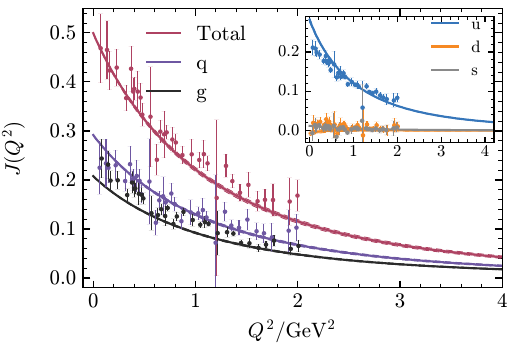}
\vspace*{4ex}

\leftline{\hspace*{0.5em}{\large{\textsf{C}}}}
\vspace*{-5ex}
\includegraphics[clip, width=0.42\textwidth]{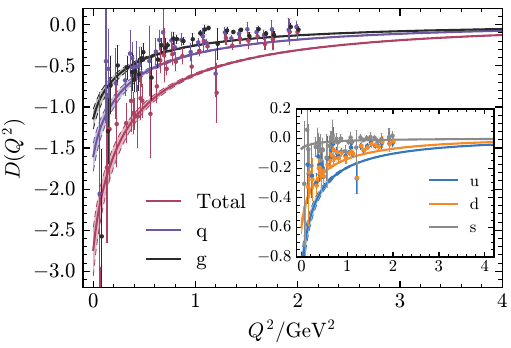}
\caption{\label{fig:GFFsat2GeV} The three gravitational form factors of the proton and their decomposition into gluon and total quark contributions(uncertainty bands as in Fig.\,\ref{fig:protonGFFs}). Inset figures show the isosinglet quark GFFs further decomposed into up-, down-, and strange-quark contributions. The total GFFs are renormalization scheme and scale independent, while all other GFFs are shown at $\zeta=2\,\rm GeV$. The lQCD points in each panel are reproduced from Ref.\,\cite{Hackett:2023rif}.}
\end{figure}

Following the AO scheme in Ref.\,\cite{Yin:2023dbw}, the results for the nucleon GFFs $A(Q^2)$ and $J(Q^2)$ at $\zeta=\zeta_2:=2\,\rm GeV$ can be obtained. According to  Sect.\,3.6\,-\,3.9 of Ref.\,\cite{Diehl:2003ny}, the AO scheme is extended to obtain the same result for $D(Q^2)$. Consequently, the species decompositions of all form factors in the gravitational current emerge and evolve in the same way. We thereby arrive at the $\zeta_2$ species decompositions shown in Fig.\,\ref{fig:GFFsat2GeV} and listed in Table\,\ref{TableZero}.

The values of $A(0)$ and $J(0)$ in Table\,\ref{TableZero} reflect the symmetry-preserving character of our analysis. While our results for the quark and gluon contributions to $A(0)$ and $J(0)$, which are consistent with the three-body approach, differ from recent lQCD results\,\cite{Hackett:2023rif}, they are in good agreement with the lQCD results extrapolated to the physical pion mass, i.e., $A^q(0)=0.618(60)$, $A^g(0)=0.427(92)$, $J^q(0)=0.285(45)$, and $J^g(0)=0.187(46)$\,\cite{Alexandrou:2020sml}. In particular, our result for $A^g(0)$ matches the value inferred from global analyses of data by many collaborations\,\cite{Hou:2019efy}: $A^{g}(0)=0.413(6)$. 

\begin{figure}[t]
\centering
  \includegraphics[clip, width=0.44\textwidth]{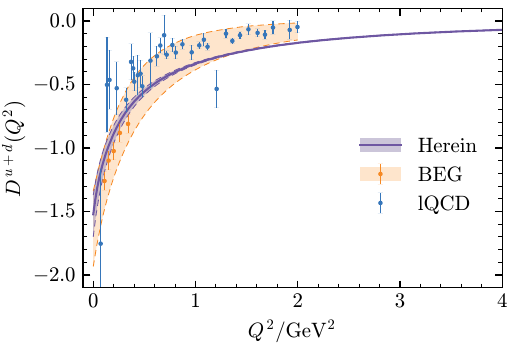}
\caption{\label{fig:DudGFF}
Result for the proton GFF $D^{u+d}(Q^2;\zeta)$ predicted herein, compared with the lQCD
results~\cite{Hackett:2023rif} and the experimentally constrained multipole parametrizations
of Ref.~\cite{Burkert:2018bqq} (BEG).
Solid purple curve within lighter purple model-uncertainty band -- prediction herein, with $\zeta = \zeta_B$;
blue points -- lQCD results, obtained with $\zeta = \zeta_2$;
orange points and associated orange band -- BEG fit from Ref.\,\cite{Burkert:2018bqq}, with $\zeta \approx \zeta_B$.}
\end{figure}

From the $D$-term GFFs in Table\,\ref{TableZero}, our study yields a nucleon $D$-term of $D(0)=-2.736^{+0.270}_{-0.319}$. Its central value is about $30\%$ smaller in magnitude compared with the result predicted by lQCD\,\cite{Hackett:2023rif}, though the two results are close when considering the combined uncertainties. This deviation may be attributed to the omission of meson cloud effects. Considering the light quarks alone, an inference from available data yields\,\cite{Burkert:2018bqq}: $D^{u+d}(0)=-1.63(29)$, which may be identified with the scale $\zeta \approx \zeta_{B}= 1.6\,\rm GeV$\,\cite{CLAS:2007clm,CLAS:2015uuo}. Our study predicts $D^{u+d}(0;\zeta_{B})=-1.52^{+0.15}_{-0.18}$, differing from the experimental results by only about $7\%$. Moreover, within uncertainties, our predictions are in $Q^2$ pointwise agreement with the extraction in Ref.\,\cite{Burkert:2018bqq} at low $Q^2$ -- see Fig.\,\ref{fig:DudGFF}. 

From the results in Table\,\ref{TableZero}, we predict that the $d/u$ ratios are not equal for the different form factors: $0.529(2)$ for $A$, $-0.024(18)$ for $J$ and $0.658(37)$ for $D$. This differs from the prediction of the three-body approach, where the $d/u$ ratio is equal to $0.529$ for all form factors. This distinction may arise because the presence of the diquark in the proton assigns different roles to the $u$ and $d$ quarks. Considering the quark and gluon contributions to $J(0)$ (proton spin), the quark contribution is greater than the gluon contribution, in agreement with that reported in Ref.\,\cite{Alexandrou:2020sml}. For the $D$-term, our result for the quark contribution is consistent with the lQCD result, but our gluon result is notably smaller. As noted above, this underestimation of the gluon contribution can likely be attributed to the omission of meson cloud effects.

Figure\,\ref{fig:GFFsat2GeV} compares our predictions for the species-separated form factors with available lQCD results. Our results for $A(Q^2)$ and $J(Q^2)$ agree with the lQCD results within uncertainties. For the form factor $D(Q^2)$, our results are harder than the lQCD results, i.e., their magnitudes decrease more slowly with increasing $Q^2$. 

\begin{figure}[t]
\leftline{\hspace*{0.5em}{\large{\textsf{A}}}}
\vspace*{-5ex}
\includegraphics[clip, width=0.42\textwidth]{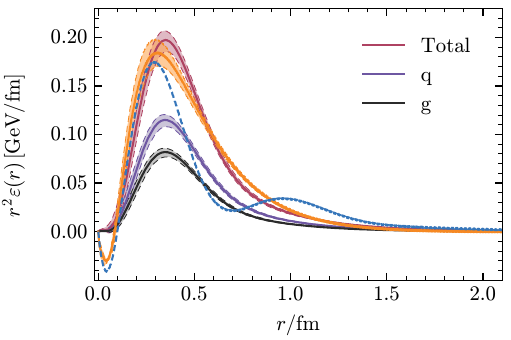}
\vspace*{4ex}

\leftline{\hspace*{0.5em}{\large{\textsf{B}}}}
\vspace*{-5ex}
\includegraphics[clip, width=0.42\textwidth]{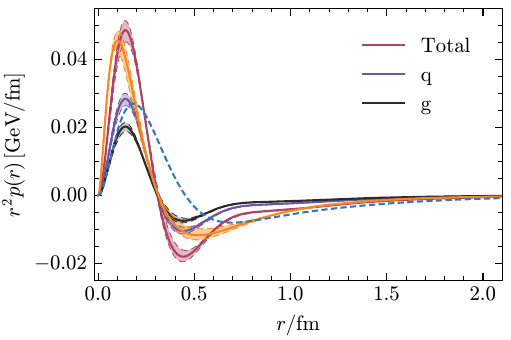}
\vspace*{4ex}

\leftline{\hspace*{0.5em}{\large{\textsf{C}}}}
\vspace*{-5ex}
\includegraphics[clip, width=0.42\textwidth]{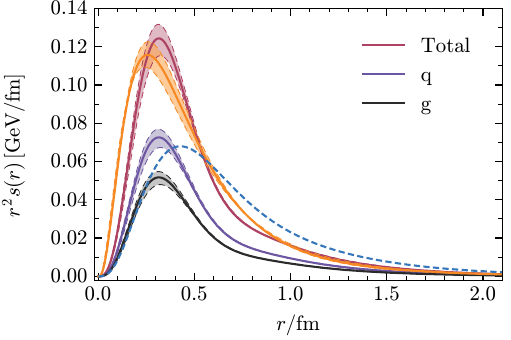}
\caption{\label{fig:densities} The quark, gluon, and total contributions to the proton energy $\epsilon(r)$ (upper), pressure $p(r)$ (middle), and shear force $s(r)$ (lower) densities in the Breit frame at $\zeta=2GeV$: solid curve--herein; dashed curve--the three-body approach. The brown solid curve represents the total result obtained via the z-expansion extrapolation (see, e.g., Appendix B of Ref.\,\cite{Cao:2025dkv}).}
\end{figure}

\section{Density profiles and radii}
\label{sec:DensityandRadii}

Using the Poincar\'e-invariant nucleon GFFs obtained herein, one can analyze various spatial densities in the proton--such as the energy, pressure, and shear force distributions--via appropriate three-dimensional Fourier transforms\,\cite{Polyakov:2018zvc}. The relevant formulas are collected in Appendix\,\ref{app:densitiesatBF}, and the physical significance of these distributions has been widely discussed\,\cite{Polyakov:2018zvc,Jaffe:2020ebz,
Freese:2021czn,Epelbaum:2022fjc,Freese:2022fat}. A stable numerical Fourier transformation requires knowledge of the complete $Q^2$-dependence of the GFFs. Since our direct numerical calculations yield stable GFF results onlt up to $Q^2 \lesssim 10\,\rm{GeV}^2$, we adopt a Pad\'e polynomial method to extrapolate the GFFs in the spacelike region (for more details, see App.\,\ref{app:densitiesatBF}). 

Figure\,\ref{fig:densities} displays the predictions for the proton energy, pressure, and shear force densities, compared with the central values obtained from three-body approach\,\cite{Yao:2024ixu}. Our predictions yield larger peak values but narrower peak widths than those of the three-body approach. Consequently, the peaks for the pressure and shear force occur at smaller radial distances $r$, yielding smaller overall radii (see Table\,\ref{Dandradii}). We note that these discrepancies are primarily driven by the differences in the $D(Q^2)$. Despite these quantitative differences, our results are consistent with the three-body\,\cite{Yao:2024ixu} and dispersion-analysis results\,\cite{Cao:2025dkv} in their overall $r$-dependence.

\begin{table}[t]
\caption{\label{Dandradii} Radii for the corresponding nucleon GFFs.(The unit of radii are fm.)}
\begin{center}\renewcommand{\arraystretch}{1.3}
\begin{tabular*}
{\hsize}{@{\extracolsep{\fill}} l c c c @{}}
\hline
  & $\sqrt{\left\langle r^2\right\rangle_{\mathrm{mass}}}$& $\sqrt{\left\langle r^2\right\rangle_{\mathrm{mech}}}$ & $\sqrt{\left\langle r^2\right\rangle_{\theta}}$  \\\hline
Herein & $0.59(1)$ & $0.51(2)$ & $0.76(1)$ \\\hline
3-body\cite{Yao:2024ixu} & $0.72(4)$ & $0.64(2)$ & $0.96(4)$ \\\hline
lQCD\cite{Hackett:2023rif}\ & $0.726(48)$ & $0.750(79)$ & -  \\\hline
DR\cite{Cao:2024zlf} & $0.70^{+0.03}_{-0.04}$ & $0.72^{+0.09}_{-0.08}$ & $0.97^{+0.03}_{-0.03}$ \\\hline
\end{tabular*}
\end{center}
\end{table}

To study the model dependence of our results on the extrapolation scheme, we also present in Fig.\,\ref{fig:densities} the results obtained from a $z$-expansion method\,\cite{Cao:2025dkv}. The comparison shows that the predictions are sensitive to the extrapolation model, particularly at short distances. Interestingly, the $z$-expansion results reproduce a peculiar feature observed in the lattice results of Ref.\,\cite{Hackett:2023rif}, namely a negative energy density in the central region. In contrast, neither the Pad\'e results obtained herein nor the $z$-expansion results from the dispersion analysis\,\cite{Cao:2025dkv} exhibit such behavior. This discrepancy suggests that the presence of a negative central energy density is highly model-dependent. Determining whether this negative energy density reflects a genuine physical phenomenon or an artifact of the extrapolation scheme will require further investigation with controlled systematic uncertainties.

\begin{figure}[t]
\includegraphics[width=\columnwidth]{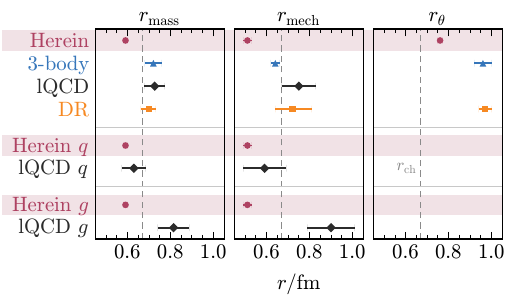}
\caption{\label{fig:radii}
Proton mass (left), mechanical (middle) and scalar trace (right) radii.
Upper group -- total radii: predictions herein (circles, shaded row) compared with
results from the three-body approach\,\cite{Yao:2024ixu} (triangles),
lQCD\,\cite{Hackett:2023rif} (diamonds), and the dispersive analysis
of Ref.\,\cite{Cao:2024zlf} (DR, squares); no lQCD result is available for $r_\theta$.
Lower groups -- quark ($q$) and gluon ($g$) radii at $\zeta_2$, predicted herein
(shaded rows) and from lQCD\,\cite{Hackett:2023rif}.
Where not visible, the uncertainty of our results is smaller than the symbol size.
Vertical dashed line -- proton charge radius, $r_{\mathrm{ch}} = 0.67\,$fm,
obtained in the same framework\,\cite{Cheng:2025yij}.
(Numerical values are listed in Table\,\ref{Dandradii}.)}
\end{figure}

It is instructive to discuss the nucleon mass, mechanical and scalar trace radii, determined by Eqs.\,\eqref{eq:radiiv2} and \eqref{eq:scalarradii}. Our results are shown in Table\,\ref{Dandradii} and Fig.\,\ref{fig:radii}. Compared with the three-body approach and lQCD results, our predicted radii are somewhat smaller. This behavior is similar to the case of the electromagnetic radius\,\cite{Cheng:2025yij} and can be naturally attributed to the omission of meson cloud effects. This is consistent with the conclusions in Ref.\,\cite{Wang:2024abv}, where the analysis indicates that meson cloud effects systematically increase the radii. Using the proton charge radius $r_{\mathrm{ch}}=0.67\,\rm fm$ calculated within the same framework\,\cite{Cheng:2025yij}, we obtain the following ratios: 
\begin{equation}
r_{\mathrm{mass}}=0.88(2) r_{\mathrm{ch}}, r_{\mathrm{mech}}=0.76(2) r_{\mathrm{ch}}, r_{\theta}=1.13(2) r_{\mathrm{ch}},    
\label{eq:radii}
\end{equation}
which yields the hierarchy 
\begin{equation}
r_{\theta} > r_{\mathrm{ch}} > r_{\mathrm{mass}} > r_{\mathrm{mech}},
\label{radiiordering}
\end{equation}
consistent with the three-body approach. Regarding the species decompositions at $\zeta_2$, we find 
\begin{equation}
\begin{aligned}
r_{\mathrm{mass}}^q&=r_{\mathrm{mass}}^g=0.59(1)\,\rm fm, \\
r_{\mathrm{mech}}^q&=r_{\mathrm{mech}}^g=0.51(2)\,\rm fm.
\label{Rmass&mech}
\end{aligned}
\end{equation}
Our quark radii are in reasonable agreement with lQCD results\,\cite{Hackett:2023rif}, which yield $0.629(56)\,\rm fm$ for the mass radius and $0.59(10)\,\rm fm$ for the mechanical radius. However, our gluon radii are notably smaller than the lQCD values of $0.814(73)\,\rm fm$ and $0.90(11)\,\rm fm$, respectively. This systematic underestimation of the gluon radii, combined with our smaller overall radii, suggests that the omitted meson cloud effects significantly affect the gluonic structure of the nucleon.

\section{Summary}
\label{sec:summary}

Using a Poincar\'e-covariant quark $+$ diquark Faddeev equation approach, we have constructed a symmetry-preserving gravitational current for the nucleon. The parameters in this current were fixed by the electromagnetic case, rendering all our predictions parameter-free. With this current, we deliver a set of predictions for the nucleon GFFs ($A$, $J$, $D$ and $\bar{c}$), including a separation of each GFF into gluon and quark pieces. 

From the results of the contribution separations according to the diquark content at the hadron scale $\zeta_H$, we observe that while the scalar diquark component dominates the nucleon's energy, spin, and mechanical properties, the axialvector component makes an indispensable contribution. Our results predict that the magnitude of $B(Q^2)$ is vanishingly small but nonvanishing at finite $Q^2$, strictly obeying the constraint $B(0) = 0$. Both the cancellation between the scalar and axial-vector diquark components and the cancellation between the u and d quarks demonstrate that the smallness of $B(Q^2)$ is deeply rooted in the nucleon’s internal structure. Furthermore, our numerical results for $\bar{c}(Q^2)$ at $\zeta_H$ confirm that our approach satisfies the fundamental symmetry requirements, with the extremely small values of $\bar{c}^{u/d}(Q^2)$ implying that these contributions play almost no role in the flavor-decomposed energy and pressure densities inside a proton. 

Our results for the proton GFFs $A(Q^2)$ and $J(Q^2)$, along with their decomposition into gluon and quark contributions, are in good agreement with recent lQCD results\,\cite{Hackett:2023rif} and the lQCD results extrapolated to the physical pion mass\,\cite{Alexandrou:2020sml}. For the GFF $D(Q^2)$, we obtain a nucleon D-term of $D(0) = -2.736^{+0.270}_{-0.319}$, which is about $30\%$ smaller in magnitude than the lQCD prediction\,\cite{Hackett:2023rif}. This underestimation primarily stems from the gluon contribution, while our quark contribution is consistent with lQCD, suggesting an omission of meson cloud effects in our framework. Meanwhile, our prediction for $D^{u+d}(Q^2;\zeta_B)$ is in agreement with recent experimental results at low $Q^2$\,\cite{Burkert:2018bqq}. Notably, the predicted $d/u$ ratios for various GFFs in our model differ from those of the three-body approach, likely because the presence of the diquark assigns different dynamical roles to the $u$ and $d$ quarks. This distinction provides a valuable probe for examining the diquark structure within the nucleons.

The omission of meson cloud effects is also reflected in the spatial radii. The nucleon mass, mechanical and scalar trace radii predicted by our GFFs are somewhat smaller than those from the three-body approach and lQCD. This deviation is primarily driven by the notably smaller gluon radii. This is consistent with the conclusions of Ref.\,\cite{Wang:2024abv}, where the analysis indicates that meson cloud effects systematically increase the radii. Together with the D-term results, our analysis suggests that meson cloud effects have a more pronounced impact on the gluon sector than on the quark sector.

Finally, our density profiles derived from two extrapolation schemes show that the predictions are sensitive to the extrapolation model, particularly at short distances. Notably, the comparison between our results and those from other methods suggests that the
presence of a negative central energy density is highly
model-dependent. Further investigation with controlled
systematic uncertainties is therefore required to ascertain
whether this negative energy density reflects a genuine
physical phenomenon or an artifact of the extrapolation scheme. Given the crucial role of the meson cloud identified in this study, a natural next step is to explicitly incorporate meson cloud effects into our framework to verify this mechanism and quantify their impact on the nucleon GFFs. 

\begin{CJK*}{UTF8}{gbsn}
\begin{acknowledgments}
We are grateful to Yang Li and Dimitra A. Pefkou for providing the GFFs data calculated via LFHQCD and lQCD, respectively. We also thank Craig D. Roberts and Hui-Yu Xing for helpful discussions and suggestions.
This work is supported by Natural Science Foundation of Anhui Province under Grant no.\ 2408085QA028 and by the National Natural Science Foundation of China under Grant No.\ 12247103.
Zhao-Qian Yao acknowledges support from Helmholtz-Zentrum Dresden-Rossendorf
through the High Potential Programme.
\end{acknowledgments}
\end{CJK*}

\appendix
\setcounter{equation}{0}
\setcounter{figure}{0}
\setcounter{table}{0}
\renewcommand{\theequation}{\Alph{section}.\arabic{equation}}
\renewcommand{\thetable}{\Alph{section}.\arabic{table}}
\renewcommand{\thefigure}{\Alph{section}.\arabic{figure}}

\section{Quark+diquark framework} 
\label{app:qdiqframework}

The model we exploit herein was introduced in Refs.\,\cite{Oettel:1998bk,Alkofer:2004yf,Cloet:2008re,Segovia:2014aza}. Combined with the material above, the information in this Appendix is sufficient for an interested practitioner to implement the framework for themselves.

\subsection{Dressed-quark propagator} 
\label{subapp:qprop}

A principal piece of the Fig.\,\ref{fig:FaddeevEquation} Faddeev kernel is the dressed-quark propagator:
\begin{equation}
\label{eq:qpropagator}
\begin{aligned}
S(p) = & -i \gamma\cdot p\, \sigma_V(p^2) + \sigma_S(p^2) \\
= & 1/[i\gamma\cdot p\,
A(p^2) + B(p^2)].    
\end{aligned}    
\end{equation}
Regarding light-quarks, the wave function renormalization and dressed-quark mass:
\begin{equation}
\label{eq:ZMdef}
Z(p^2)=1/A(p^2)\,,\;M(p^2)=B(p^2)/A(p^2)\,,
\end{equation}
respectively, receive material momentum-dependent corrections at infrared momenta\,\cite{Binosi:2016wcx}: $Z(p^2)$ is suppressed and $M(p^2)$ enhanced. These features are an expression of emergent hadron mass (EHM)\,\cite{Roberts:2021nhw, Binosi:2022djx, Ding:2022ows}.

The following parametrisation of $S(p)$, which exhibits the characteristics described above, has long been used in hadron studies\,\cite{Cloet:2008re,Segovia:2014aza,Chen:2018nsg,Lu:2019bjs,Chen:2021guo,Chen:2023zhh,Cheng:2025yij}:
\begin{subequations}
\label{eq:ssmsvm}
\begin{align}
\label{eq:ssm} \bar\sigma_S(x) =&  2\,\bar m \,{\cal F}(2 (x+\bar m^2)) \nonumber \\
&+ {\cal F}(b_1 x) \,{\cal F}(b_3 x) \,
\left[b_0 + b_2 {\cal F}(\epsilon x)\right]\,,\\ 
\label{eq:svm} \bar\sigma_V(x) = & \frac{1}{x+\bar m^2}\, \left[ 1 - {\cal F}(2
(x+\bar m^2))\right]\,,
\end{align}
\end{subequations}
with $x=p^2/\lambda^2$, $\bar m$ = $m/\lambda$,
\begin{equation}
\label{eq:calF}
{\cal F}(x)= \frac{1-\mbox{\rm e}^{-x}}{x}  \,,
\end{equation}
$\bar\sigma_S(x) = \lambda\,\sigma_S(p^2)$ and $\bar\sigma_V(x) =
\lambda^2\,\sigma_V(p^2)$. The mass scale, $\lambda=0.566\,$GeV, and parameter values
\begin{equation}
\label{eq:tableA}
\begin{array}{ccccc}
   \bar m& b_0 & b_1 & b_2 & b_3 \\\hline
   0.00897 & 0.131 & 2.90 & 0.603 & 0.185
\end{array}\;,
\end{equation}
associated with Eqs.\,\eqref{eq:ssmsvm} were fixed in analyses of light-meson observables\,\cite{Burden:1995ve,Hecht:2000xa}\footnote{$\epsilon=10^{-4}$ in Eq.\,\eqref{eq:ssm} acts only to decouple the large- and intermediate-$p^2$ domains.}. The expressions in Eqs.\,\eqref{eq:ssmsvm} guarantee dressed-quark confinement via the violation of reflection
positivity; see, e.g., Sect.\,5 of Ref.\,\cite{Ding:2022ows}.

The dimensionless $u = d$ current-quark mass $\bar{m}$ in Eq.\,\eqref{eq:tableA} corresponds to $m_q=5.08\,{\rm MeV}$. The propagator then yields a Euclidean constituent-quark mass $\label{MEq} M_q^E = 0.33\,{\rm GeV}$, defined as the solution of $p^2=M^2(p^2)$. The ratio $M_q^E/m_q = 65$ is one expression of dynamical chiral symmetry breaking (DCSB), a corollary of EHM, in the parametrization of $S(p)$. This ratio highlights the infrared enhancement of the dressed-quark mass function.

The dressed-quark mass function produced by Eqs.\,\eqref{eq:ssmsvm}--\eqref{eq:tableA} is drawn in Fig.~13 of Ref.\,\cite{Chen:2021guo}. Although the algebraic representation is simple and was introduced long ago\,\cite{Burden:1995ve, Hecht:2000xa}, before practicable numerical solutions of QCD’s gap equation were commonly obtained, it remains a sound representation of contemporary numerical results, as confirmed by the good agreement shown in Fig.~13 of Ref.\,\cite{Chen:2021guo}.

\subsection{Correlation amplitudes} 
\label{subapp:diqAmp}

Following almost forty years of study, it is now clear that a realistic description of the nucleon can be achieved only if one includes the two positive-parity diquark correlations, namely the isoscalar-scalar ($0^+$) and isovector-axialvector ($1^+$) diquarks. Negative-parity diquarks play practically no role in the positive parity nucleon\,\cite{Barabanov:2020jvn}. Baryon structure models that ignore axialvector diquarks are incompatible with existing data; see, e.g., Refs.\,\cite{Chen:2021guo, Cui:2021gzg, Chen:2023zhh}.

For $0^+$, $1^+$ diquarks, the dominant structures in their correlation amplitudes are:
{\allowdisplaybreaks
\begin{subequations}
\label{eq:diqAmplitude}
\begin{eqnarray}
\label{Gamma0p} \Gamma^{0^+}(k;K) &= g_{0^+} \gamma_5 C t^0_f \vec{H}_c {\cal F}(k^2/\omega_{0^+}^2) \,, \\
\label{Gamma1p} \vec{\Gamma}^{1^+}_\mu (k;K) &= g_{1^+} i \gamma_\mu C \vec{t}_f \vec{H}_c {\cal F}(k^2/\omega_{1^+}^2)\,.
\end{eqnarray}
\end{subequations}
Here $K$ is the total momentum of the correlation, $k$ is a two-body relative momentum, $\cal F$ is the function in Eq.\,\eqref{eq:calF}, $\omega_{J^P}$ is a size parameter, and $g_{J^P}$ is a channel coupling fixed by normalization - see Eq.\,\eqref{eq:Amplitudenorm} below; $C=\gamma_2 \gamma_4$ is the charge-conjugation matrix; $t^0$ and $\vec{t}=(t^1,t^2,t^3)$ introduce the flavor structure: 
\begin{equation}
\begin{aligned}
& t_f^0=\frac{i}{\sqrt{2} } \tau^2,\ t_f^1=\frac{1}{2}\left(\tau^0+\tau^3\right), \\
& t_f^2=\frac{1}{\sqrt{2} } \tau^1,\ t_f^3=\frac{1}{2}\left(\tau^0-\tau^3\right),
\end{aligned}
\label{eq:diqflavour}
\end{equation}
with $\tau^0=\rm{diag}[1,1]$, $\{\tau^j,j=1,2,3\}$ being the Pauli matrices; and $\vec{H}_c=\{i \lambda_c^7, -i \lambda_c^5, i \lambda_c^2\}$, with $\{\lambda_c^k, k=1,...,8\}$ being the Gell-Mann matrices in color space, expresses the color-antitriplet character of the diquarks. 

The amplitudes in Eqs.\,\eqref{eq:diqAmplitude} are canonically normalized: 
\begin{subequations}
\label{eq:Amplitudenorm}
\begin{align}
2 \,K_\mu = & \left[ \frac{\partial}{\partial Q_\mu} \operatorname{Tr}_{\text{CDF}} \int \frac{d^4 q}{(2\pi)^4}\,  \nonumber \right. \\
 & \times \left. \bar\Gamma^{0^+}(q;-K)  \, S(q+Q/2)\Gamma^{0^+}(q;K) \, \nonumber \right. \\
 & \times \left. S^{\rm T}(-q+Q/2)  \right]_{Q=K}^{{K^2=-m_{0^+}^2}},\\
2 \,K_\mu T_{\alpha \beta}(K) = &
\left[ \frac{\partial}{\partial Q_\mu} \operatorname{Tr}_{\text{CDF}} \int \frac{d^4 q}{(2\pi)^4}\,  \nonumber \right. \\
& \times \left. \bar\Gamma^{1^+}_\alpha(q;-K) \, S(q+Q/2) \, \Gamma^{1^+}_\beta(q;K) \, \nonumber \right. \\
 & \times \left. S^{\rm T}(-q+Q/2)  \right]_{Q=K}^{{K^2=-m_{1^+}^2}},
\end{align}
\end{subequations}
where $\bar\Gamma(k;K) = C^\dagger \Gamma(-k;K)^{\rm T} C$, with $(\cdot)^{\rm T}$ denoting matrix transpose. The coupling strength in each channel, $g_{J^P}$ in Eq.\,\eqref{eq:diqAmplitude}, is fixed by the associated value of $\omega_{J^P}$ via Eqs.\,\eqref{eq:Amplitudenorm}.
}


\subsection{Diquark propagators and couplings} 
\label{subapp:diqprop}

A propagator is associated with each quark+quark correlation in Fig.\,\ref{fig:FaddeevEquation}\,\cite{Segovia:2014aza}:
{\allowdisplaybreaks
\begin{subequations}
\label{eq:diqprop}
\begin{align}
\Delta^{0^+}(K) & = \Pi^{0^+}(K^2)\,,\\
%
\Delta^{1^+}_{\mu\nu}(K) & = \left[ \delta_{\mu\nu} + \frac{K_\mu K_\nu}{m_{1^+}^2} \right] \Pi^{1^+}(K^2)\,,
\end{align}
\end{subequations}
with
\begin{equation}
\label{eq:defofPiJP}
 \Pi^{J^P}(K^2) = \frac{1}{m_{J^P}^2} \, {\mathpzc F}(K^2/\omega_{J^P}^2)\,.
\end{equation}
These forms ensure that each diquark is confined within a baryon: although the propagators are free-particle-like at spacelike momenta, they are pole-free on the timelike axis. This suffices to ensure confinement through the violation of reflection positivity.

The following values are commonly used for the masses in Eq.\,\eqref{eq:diqprop} \cite{Chen:2017pse,Chen:2021guo,Liu:2022ndb,Liu:2022nku}:
\begin{subequations}
\label{eq:diqmasses}
\begin{align}
m_{[ud]_{0^+}} &= 0.79\,\mbox{GeV}\,,\\
m_{\{uu\}_{1^+}} = m_{\{ud\}_{1^+}} &= m_{\{dd\}_{1^+}} = 0.89\,\mbox{GeV}\,.
\end{align}
\end{subequations}
The widths in Eq.\,\eqref{eq:diqprop} are related to these masses via\,\cite{Segovia:2014aza}
\begin{equation}
\label{eq:massRomega}
\omega_{J^P}^2=\frac{1}{2} m_{J^P}^2.
\end{equation}
This correspondence reflects the free-particle-like propagation of diquark correlations within the nucleon and also reduces the number of parameters.

Using Eq.\,\eqref{eq:massRomega} and Eqs.\,\eqref{eq:diqAmplitude}, \eqref{eq:Amplitudenorm}, \eqref{eq:diqmasses}, one obtains 
\begin{equation}
g_{0^+} = 14.94\,,\,\,\,\,\,
g_{1^+} = 12.98\,.
\end{equation}
Since it is the coupling-squared which appears in the Faddeev kernels, $0^+$ diquarks dominate the Faddeev amplitudes of $J=1/2^+$ baryons, while $1^+$ diquarks also play a measurable role because $g^2_{1^+}/g^2_{0^+} \simeq 0.75$. These outcomes are visible in, e.g., Fig. 2 of Ref.\,\cite{Liu:2022ndb}.
}


\subsection{Nucleon Faddeev amplitudes}
\label{subapp:NAmplitude}
Having defined all elements of the kernel, one can solve the Faddeev equation ( Fig.\,\ref{fig:FaddeevEquation}) to obtain both the nucleon mass squared, $m_N^2$, and bound-state amplitude. As for all baryons, the nucleon is described by the sum of three amplitudes
{\allowdisplaybreaks
\begin{equation}
\Psi^N = \psi^N_1 + \psi^N_2 + \psi^N_3\,,    
\end{equation}
where the subscript identifies the bystander quark, i.e., the quark that is not participating in a diquark correlation. $\psi^N_{1,2}$ are obtained from $\psi^N_3=:\psi^N$ by cyclic permutations of all quark labels.

For the nucleon,
\begin{align}
\nonumber
& \psi^N(p_i,\alpha_i,\sigma_i) \\
\nonumber
 = & [\Gamma^{0^+}(k;K)]^{\alpha_1 \alpha_2}_{\sigma_1 \sigma_2} \, \Delta^{0^+}(K) \,[\Psi^{0^+}(\ell;P) u(P)]^{\alpha_3}_{\sigma_3} \\
 + & [{\Gamma}^{j 1^+}_\mu(k;K)]^{\alpha_1 \alpha_2}_{\sigma_1 \sigma_2} \, \Delta^{1^+ }_{\mu\nu}(K) \, [{\Psi}_{\nu }^{j 1^+}(\ell;P) u(P)]^{\alpha_3}_{\sigma_3} \,,
\label{eq:FaddeevAmp}
\end{align}
where $(p_i,\alpha_i,\sigma_i)$ are the momentum, spin and isospin labels of the quarks comprising the bound state; $P=p_1 + p_2 + p_3=p_d+p_q$ is the baryon's total momentum; $k=(p_1-p_2)/2$, $K=p_1+p_2=p_d$, $\ell = (-K + 2 p_3)/3$; the $j$ sum runs over the axialvector diquarks in the given bound state (Eq.\,\eqref{eq:diqflavour}): 1, 2 for proton and 2, 3 for neutron; and $u(P)$ is a Euclidean spinor, as well as a spinor in isospin space, with $\phi_+=\text{col}(1,0)$ for the proton and $\phi_-=\text{col}(0,1)$ for the neutron (see Appendix\,B of Ref.\,\cite{Segovia:2014aza} for details).

The remaining terms in Eq.\,\eqref{eq:FaddeevAmp} are the following matrix-valued functions:
\begin{subequations}
\label{eq:AplitudeBasis}
\begin{align}
\Psi^{0^+}(\ell;P) & = \sum_{k=1}^2 {\mathpzc s}_{k}(\ell^2,\ell\cdot P)\,  {\mathpzc S}^k(\ell;P)
\, \frac{\lambda^0_c}{\sqrt{3}}
\, s^0_f \,, \\
{\Psi}_{\nu}^{j 1^+}(\ell;P)  & = \sum_{k=1}^6 {\mathpzc a}^{j}_{k}(\ell^2,\ell\cdot P)\, \gamma_5 {\mathpzc A}^k_\nu(\ell;P)
\,  \frac{\lambda^0_c}{\sqrt{3}}
\, s^{j}_f\,,
\end{align}
\end{subequations}
where
\begin{align}
\label{eq:diracbasis}
\nonumber
{\mathpzc S}^1 & = {\mathbf I}_{\rm D} \,,\;
{\mathpzc S}^2  = i \gamma\cdot\hat\ell - \hat\ell \cdot\hat P {\mathbf I}_{\rm D}\,, \\
{\mathpzc A}^1_\nu & =  \gamma\cdot \hat\ell^\perp \hat P_\nu\,,\;
{\mathpzc A}^2_\nu  = - i \hat P_\nu {\mathbf I}_{\rm D}\,,\;
{\mathpzc A}^3_\nu  = \gamma\cdot\hat\ell^\perp \hat\ell^\perp_\nu\,, \\
\nonumber
{\mathpzc A}^4_\nu & = i\hat \ell_\nu^\perp {\mathbf I}_{\rm D}\,,\;
{\mathpzc A}^5_\nu = \gamma_\nu^\perp - {\mathpzc A}^3_\nu\,,\;
{\mathpzc A}^6_\nu = i \gamma_\nu^\perp \gamma\cdot\hat\ell^\perp - {\mathpzc A}^4_\nu\,,
\end{align}
are the Dirac basis matrices, with $({\mathbf I}_{\rm D})_{rs}=\delta_{rs}$, $\hat\ell^2=1$, $\hat P^2 = -1$, $\ell_\nu^\perp = \hat\ell_\nu +\hat\ell\cdot\hat P \hat P_\nu$, $\gamma_\nu^\perp = \gamma_\nu +\gamma\cdot\hat P \hat P_\nu$ and $\hat\ell^\perp \cdot \hat\ell^\perp=1$; $\lambda^0_c/\sqrt{3}$ is the color matrix with $\lambda^0_c={\rm diag}[1,1,1]$; and $s^0_f$ and $s^j_f$ are the flavor matrices of the quark+diquark amplitudes, which are obtained by removing the diquarks' flavor matrices, Eq.\,\eqref{eq:diqflavour}, from the nucleon's full amplitude:
\begin{subequations}
\label{eq:qfmat}
\begin{align}
s_f^0 &= \tau^0\,,\\
s_f^1 &= \tfrac{1}{\surd 6}(\tau^1-i\tau^2)\,,\\
s_f^2 &= -\tfrac{1}{\surd 3}\tau^3\,,\\
s_f^3 &= -\tfrac{1}{\surd 6}(\tau^1+i\tau^2)\,.
\end{align}
\end{subequations}

In solving the Faddeev equation, one projects the equation onto isospin $+1/2$ or $-1/2$ for the proton or neutron, respectively, and then employs a positive-energy projection operator,
\begin{equation}
\Lambda_{+}(P):=\frac{1}{2 m_N} \sum_{s= \pm} u(P, s) \bar{u}(P, s)=\frac{-i \gamma \cdot P+m_N}{2 m_N},    
\end{equation}
to obtain an integral equation for the eight scalar
functions $\{{\mathpzc s}_{1,2}(\ell^2,\ell\cdot P)\}$, $\{{\mathpzc a}_{1,...,6}(\ell^2,\ell\cdot P)\}$ that describe the relative quark + diquark momentum correlations inside the nucleon. Isospin symmetry entails, e.g., $\sqrt{2} {\mathpzc a}_k^2=-{\mathpzc a}_k^1$. Details are provided in Sect. 2 of Ref.\,\cite{Alkofer:2004yf}.}

\section{Color and Flavor coefficients} 
\label{app:CFcoefficients}

The nucleon's Faddeev equation and the current diagrams of Sect.\,\ref{subsec:NGVertex} require appropriate color and flavor coefficients. These are obtained by projecting the color and flavor matrices of the diquark amplitudes, Eqs.\,\eqref{eq:diqflavour}, and the quark + diquark amplitudes, Eqs.\,\eqref{eq:qfmat}, onto the proton isospinor $\phi_+=\rm col(1,0)$ or the neutron isospinor $\phi_-=\rm col(0,1)$.

For the nucleon's Faddeev equation, the associated color-flavor coefficients have been obtained in Refs.\,\cite{Cloet:2008re,Segovia:2014aza,Chen:2021guo}. For the form-factor diagrams of Fig.\,\ref{fig:NGVertex}, their flavor
factors are obtained using Eqs.\,\eqref{eq:diqflavour} and \eqref{eq:qfmat}. We have
\begin{equation}
\label{eq:cfsofDia1}
\phi_{+(-)}^{\rm T}\left(s_f^m\right)^{\rm T}\left(\tau^a\right)\left(s_f^m\right)\phi_{+(-)},
\end{equation}
for the probe-quark diagram(Diagram 1),
\begin{equation}
\label{eq:cfsofDia2&3}
\phi_{+(-)}^{\rm T}\left(s_f^m\right)^{\rm T}\left(s_f^n\right) \phi_{+(-)}\operatorname{Tr}_{\mathrm{F}} \left[\left(t_f^m\right)^{\rm T}\left(t_f^n\right)\left(\tau^a\right)^{\rm T}\right],
\end{equation}
for the probe-diquark diagrams (Diagram 2 and 3), and 
\begin{equation}
\label{eq:cfsofDia4}
\phi_{+(-)}^{\rm T}\left(s_f^m\right)^{\rm T}(t_f^n)\left(\tau^a\right)^{\rm T}(t_f^m)^{\rm T}\left(s_f^n\right)\phi_{+(-)},
\end{equation}
for the exchange diagram (Diagram 4). The seagull case is slightly more complicated: we need to treat the bystander and exchange quarks’ legs separately.
From the seagull diagram (Diagram 5), we have
\begin{equation}
\label{eq:cfsofDia5ex}
\phi_{+(-)}^{\rm T}\left(s_f^m\right)^{\rm T}(t_f^n)\left(\tau^a\right)^{\rm T}(t_f^m)^{\rm T}\left(s_f^n\right)\phi_{+(-)},
\end{equation}
for the exchange leg, and 
\begin{equation}
\label{eq:cfsofDia5by}
\phi_{+(-)}^{\rm T}\left(s_f^m\right)^{\rm T}\left(\tau^a\right)(t_f^n)(t_f^m)^{\rm T}\left(s_f^n\right)\phi_{+(-)},
\end{equation}
for the bystander leg. From the conjugate seagull diagram (Diagram 6), we have
\begin{equation}
\label{eq:cfsofDia6ex}
\phi_{+(-)}^{\rm T}\left(s_f^m\right)^{\rm T}(t_f^n)\left(\tau^a\right)^{\rm T}(t_f^m)^{\rm T}\left(s_f^n\right)\phi_{+(-)},
\end{equation}
for the exchange leg, and
\begin{equation}
\label{eq:cfsofDia6by}
\phi_{+(-)}^{\rm T}\left(s_f^m\right)^{\rm T}(t_f^n)(t_f^m)^{\rm T}\left(\tau^a\right)\left(s_f^n\right)\phi_{+(-)},
\end{equation}
for the bystander leg. The values of $m(n)$ are $0,1,2,3$. 
The color factors of these diagrams are the same as in the
electromagnetic case: for the impulse-approximation diagrams (Diagrams 1–3), the color coefficients are “$1$”, and for the exchange or seagull diagrams (Diagrams 4–6), the color coefficients are “$-1$”.

\section{Breit frame density profiles and radii} 
\label{app:densitiesatBF}

The Breit-frame energy $\epsilon^a(r)$, pressure $p^a(r)$, and shear force $s^a(r)$ distributions are defined via the following formulae\,\cite{Polyakov:2018zvc} ($t=Q^2$): 
\begin{subequations}
\begin{align}
& \varepsilon^a(r) =m_N\left[A^a(t)+\frac{t\left[D^a(t)+A^a(t)-2 J^a(t)\right]}{4 m_N^2}\right]_{\mathrm{FT}}, \\
& p^a(r) =\frac{1}{6 m_N} \frac{1}{r^2} \frac{d}{d r} r^2 \frac{d}{d r}\left[D^a(t)\right]_{\mathrm{FT}}, \\
& s^a(r) =-\frac{1}{4 m_N} r \frac{d}{d r} \frac{1}{r} \frac{d}{d r}\left[D^a(t)\right]_{\mathrm{FT}},
\end{align}
\label{eq:densities}
\end{subequations}
which involve the Fourier transform 
\begin{equation}
\left[f(t)\right]_{\mathrm{FT}}=\int \frac{d^3 q}{(2 \pi)^3} \mathrm{e}^{i \vec{q} \cdot \vec{r}} f\left(t \rightarrow|\vec{q}|^2\right),
\label{eq:FT}
\end{equation}
\begin{table}[t!]
\caption{\label{fitParameters} Parameters in the Pad\'e extrapolation, cf. Eq.\,\eqref{eq:Padefitting}, obtained by fitting to the numerical results of the GFFs. ``$-$" indicates that it is set to zero during the fitting process.(The uncertainty arising from the variation of the diquark mass is not included here.)}
\begin{center}
\begin{tabular*}
{\hsize}
{
c@{\extracolsep{0ptplus1fil}}
|c@{\extracolsep{0ptplus1fil}}
c@{\extracolsep{0ptplus1fil}}
c@{\extracolsep{0ptplus1fil}}
c@{\extracolsep{0ptplus1fil}}
c@{\extracolsep{0ptplus1fil}}}\hline
GFF  & $a_0$ & $a_1$ & $b_1$ & $b_2$ & $b_3$ \\
\hline
$A^{\rm Total}$ & $1.000$ & $0.125$ & $1.127$ & $0.253$ & $0.087$ \\
\hline
$A^q$ & $0.584$ & $0.073$ & $1.127$ & $0.253$ & $0.087$ \\
\hline
$A^g$ & $0.416$ & $0.052$ & $1.127$ & $0.253$ & $0.087$ \\
\hline
$J^{\rm Total}$ & $0.500$ & $0.072$ & $1.101$ & $0.362$ & $0.115$ \\
\hline
$J^q$ & $0.292$ & $0.042$ & $1.101$ & $0.362$ & $0.115$ \\
\hline
$J^g$ & $0.208$ & $0.030$ & $1.101$ & $0.362$ & $0.115$ \\
\hline
$-D^{\rm Total}$ & $2.736$ & $-$ & $3.543$ & $0$ & $0.088$ \\
\hline
$-D^q$ & $1.598$ & $-$ & $3.543$ & $0$ & $0.088$ \\
\hline
$-D^g$ & $1.138$ & $-$ & $3.543$ & $0$ & $0.088$ \\
\hline
\end{tabular*}
\end{center}
\end{table}
with $r=|\vec{r}|$ and $a \in\{q, g, \rm Total\}$. (The
quark and gluon contributions to the pressure and energy
densities additionally depend on the GFF $\bar{c}^a(t)$. 
Here, we neglect this contribution since the magnitudes of $\bar{c}^{q/g}(t)$ we predict are extremely small and $\bar{c}^{\rm Total}(t)=0$.) 

The root-mean-square radii of the energy density and the longitudinal force density
\begin{equation}
F^a_{\|}(r)=p^a(r)+2 s^a(r) / 3
\label{eq:Lforce}
\end{equation}
yield the mass and mechanical radii of the proton\,\cite{Polyakov:2018zvc}:
\begin{subequations}
\begin{align}
&\left\langle r_a^2\right\rangle_{\mathrm{mass}}=\frac{\int d^3 r\,r^2\,\varepsilon^a(r)}{\int d^3 r\,\varepsilon^a(r)}, \\
&\left\langle r_a^2\right\rangle_{\mathrm{mech}}=\frac{\int d^3 r\,r^2\,F^a_{\|}(r)}{\int d^3 r\,F^a_{\|}(r)}.
\end{align}
\label{eq:radiiv1}   
\end{subequations}
In a common interpretation, the mechanical radius measures
the extent of the nucleon’s normal force distribution. These
expressions are equivalent to the following forms, which are typically easier to use:
\begin{subequations}
\begin{align}
& \left\langle r_a^2\right\rangle_{\mathrm{mass}}=\frac{-6 A^{a\prime}(0)-\frac{3}{2 m_N^2} [A^a(0)-2J^a(0)+D^a(0)]}{A^a(0)}, \label{eq:rofmass}\\
& \left\langle r_a^2\right\rangle_{\mathrm{mech}}=\frac{6}{\int_0^{\infty} d t[D^a(t) / D^a(0)]} \label{eq:rofmech}.
\end{align}
\label{eq:radiiv2}    
\end{subequations}
Regarding Eq.\,\eqref{eq:rofmech}, it is evident that the mechanical
radius is determined solely by the integral of the $Q^2=0$
unit-normalized version of the nucleon D-term function. In addition, the radius of of trace anomaly(also called scalar trace) has also attracted attention\,\cite{Cao:2024zlf, Polyakov:2018zvc}:
\begin{equation}
\left\langle r_a^2\right\rangle_{\theta}=\frac{-6 A^{a\prime}(0)-\frac{3}{2 m_N^2} [A^a(0)-2J^a(0)+3 D^a(0)]}{A^a(0)}.
\label{eq:scalarradii}  
\end{equation}

Our numerical calculations can only yield stable results of GFFs for $Q^2 \lesssim 10 \rm{GeV}^2$. However, to obtain the spatial distributions, we must extrapolate the GFFs up to $Q^2 \to +\infty$ while maintaining reasonable asymptotic behavior. The large $Q^2$ behavior of the spacelike nucleon GFFs is determined by the QCD hard scattering mechanism, up to logarithmic corrections, in Refs.\,\cite{Tong:2021ctu,Tong:2022zax} to be $A\left(Q^2\right) \sim J\left(Q^2\right) \sim 1 / Q^4$ and $D\left(Q^2\right) \sim 1 / Q^6$. With the above asymptotic behavior, we perform a Pad\'e polynomial extrapolation for the nucleon GFFs:
\begin{equation}
F(t=Q^2)=\frac{a_0+a_1 t}{1+b_1 t+b_2 t^2+b_3 t^3}.
\label{eq:Padefitting}
\end{equation}
For the GFF $D(Q^2)$, to ensure the correct asymptotic behavior, we set $a_1=0$. The corresponding fit parameters are listed in Table\,\ref{fitParameters}.


\bibliography{CollectedBiB}

\end{document}